\documentclass[fleqn,usenatbib]{mnras}

\usepackage[T1]{fontenc}

\DeclareRobustCommand{\VAN}[3]{#2}
\let\VANthebibliography\thebibliography
\def\thebibliography{\DeclareRobustCommand{\VAN}[3]{##3}\VANthebibliography}

\usepackage{graphicx}	
\usepackage{amsmath}	
\usepackage{amssymb}	

\usepackage{newtxtext,newtxmath}

\newcommand{\Msun}{\,\rmn{M}_{\sun}}

\newcommand{\Gyr}{\,\rmn{Gyr}}
\newcommand{\Myr}{\,\rmn{Myr}}

\newcommand{\kpc}{\,\rmn{kpc}}
\newcommand{\Mpc}{\,\rmn{Mpc}}

\newcommand{\Mag}{\,\rmn{mag}}
\newcommand{\Arcsec}{\,\rmn{arcsec}}

\newcommand{\kappaco}{\kappa_\mathrm{co}}

\title[Disc profile breaks in EAGLE]{Using the EAGLE simulations to elucidate the origin of disc surface brightness profile breaks as a function of mass and environment}

\author[J. Pfeffer et al.]{Joel L. Pfeffer,$^{1}$\thanks{E-mail: joel.pfeffer@icrar.org}
Kenji Bekki,$^{1}$
Duncan A. Forbes,$^{2}$
Warrick J. Couch,$^{2}$
B\"{a}rbel S. Koribalski$^{3,4}$
\\
$^{1}$International Centre for Radio Astronomy Research (ICRAR), M468, University of Western Australia, 35 Stirling Hwy, Crawley, WA 6009, Australia \\
$^{2}$Centre for Astrophysics \& Supercomputing, Swinburne University of Technology, Hawthorn VIC 3122, Australia \\
$^3$Australia Telescope National Facility, CSIRO Astronomy and Space Science, P.O. Box 76, Epping, NSW 1710, Australia \\
$^4$School of Science, Western Sydney University, Locked Bag 1797, Penrith, NSW 2751, Australia 
}

\date{Accepted XXX. Received YYY; in original form ZZZ}

\pubyear{2021}

\begin{document}
\label{firstpage}
\pagerange{\pageref{firstpage}--\pageref{lastpage}}
\maketitle

\begin{abstract}
We analyse the surface brightness profiles of disc-type galaxies in the EAGLE simulations in order to investigate the effects of galaxy mass and environment on galaxy profile types.
Following observational works, we classify the simulated galaxies by their disc surface brightness profiles into single exponential (Type~I), truncated (Type~II) and anti-truncated (Type~III) profiles.
In agreement with previous observation and theoretical work, we find that Type~II discs result from truncated star-forming discs that drive radial gradients in the stellar populations.
In contrast, Type~III profiles result from galaxy mergers, extended star-forming discs or the late formation of a steeper, inner disc.
We find that the EAGLE simulations qualitatively reproduce the observed trends found between profile type frequency and galaxy mass, morphology and environment, such as the fraction of Type~III galaxies increasing with galaxy mass, and the the fraction of Type~II galaxies increasing with Hubble type.
We investigate the lower incidence of Type~II galaxies in galaxy clusters, finding, in a striking similarity to observed galaxies, that almost no S0-like galaxies in clusters have Type~II profiles. Similarly, the fraction of Type~II profiles for disc-dominated galaxies in clusters is significantly decreased relative to field galaxies.
This difference between field and cluster galaxies is driven by star formation quenching. Following the cessation of star formation upon entering a galaxy cluster, the young stellar populations of Type II galaxies simply fade, leaving behind Type I galaxies.
\end{abstract}




\section{Introduction} \label{sec:intro}

The structural properties of galaxies are intimately linked to their formation and assembly. 
Originally thought to have exponential profiles \citep{Patterson_40, deVaucouleurs_59}, the stellar surface brightness profiles of disc galaxies have now been shown to occur in three main types: single exponential (Type~I), truncated (down-bending, Type~II) and anti-truncated (up-bending, Type~III) profiles \citep{Freeman_70, vanDerKruit_79, Erwin_et_al_05, Pohlen_and_Trujillo_06}.

Though the exact fractions of disc types differ between studies \citep[possibly due to differing galaxy samples or methodology, c.f.][]{Mendez-Abreu_et_al_17}, some general trends with galaxy mass and morphology are clear.
With increasing galaxy mass, the fraction of Type~I profiles decreases and the fraction of Type~III profiles increases \citep{Laine_et_al_16, Tang_et_al_20}.
The frequency of Type~II galaxies increases with Hubble type (i.e. are more frequent in disc-dominated galaxies), while Type~III profiles are more common in early-type galaxies \citep{Pohlen_and_Trujillo_06, Gutierrez_et_al_11, Laine_et_al_16, Mendez-Abreu_et_al_17, Tang_et_al_20}.
Approximately $8$ per cent of galaxies also have composite Type~II+III profiles \citep{Gutierrez_et_al_11}.

In Type~II (truncated) galaxies, the break in the exponential light profile has been observed to coincide with a minimum in the colour, age and mass-to-light ratio profiles \citep{Azzollini_et_al_08a, Bakos_et_al_08, Zheng_et_al_15, Ruiz-Lara_et_al_16}. The breaks are largely absent in the mass profiles of Type~II galaxies \citep{Bakos_et_al_08, Tang_et_al_20} and are thus related to radial gradients in the stellar populations \citep[similarly found in the simulations of][]{Sanchez-Blazquez_et_al_09}.
The disc breaks may result from star formation thresholds in the outer parts of galaxies \cite[e.g.][]{Kennicutt_89, Schaye_04}, and simulations of disc galaxies have found that the break may be a result of a star formation cutoff associated with a drop in cold gas density \citep{Roskar_et_al_08, Martinez-Serrano_et_al_09, Sanchez-Blazquez_et_al_09}. 
This picture is consistent with the down-bending star-formation rate (SFR) profiles of Type~II galaxies \citep{Tang_et_al_20}.
The radial position of the break has also been found to increase with cosmic time \citep{Trujillo_and_Pohlen_05, Azzollini_et_al_08b} and galaxy mass \citep{Pohlen_and_Trujillo_06, Munoz-Mateos_et_al_13}, consistent with inside-out growth of disc galaxies.

A number of origins have been suggested for Type~III (anti-truncated) galaxies. 
They may be the result of galaxy mergers \citep[e.g.][]{Bekki_98, Bournaud_et_al_05}, other tidal disturbances or galaxy harassment \citep{Watkins_et_al_19}, star formation from accreted gas \citep{Roediger_et_al_12, Wang_et_al_18}, stellar migration induced by a long-lived bar \citep{Herpich_et_al_17} or the combination of a pseudo-bulge and an outer Type~I disc \citep{Silchenko_et_al_18}.
This diversity of origins may be reflected in the structure of the galaxies, where some anti-truncations may be the result of a disc embedded within an outer spheroid, while others are a continuation of the disc \citep{Erwin_et_al_05}.
Type~III galaxies have been found to have up-bending SFR profiles, though without strong radial gradients in the stellar populations \citep[i.e. the surface brightness and mass profiles are closely related in Type~III galaxies,][]{Tang_et_al_20}.
\citet{Borlaff_et_al_18} found Type~III break radius scaling relations are similar from $z=0.6$ to $z=0$, indicating a `stable' formation process.

A curious and so far, unexplained, observation is that, despite being common for field S0 galaxies, Type~II profiles are completely absent for S0 galaxies in the Virgo cluster \citep{Erwin_et_al_12}. 
\citet{Roediger_et_al_12} and \citet{Raj_et_al_19} similarly found that the fraction of Type~II profiles in Virgo and Fornax cluster disc galaxies are suppressed relative to field galaxies (which decrease from 50-60 per cent in field galaxies to 34 per cent in the Virgo cluster and 38 per cent in the Fornax cluster, respectively).
\citet{Lee_et_al_18} also found a similar environmental dependence for Type~II profiles of dwarf galaxies in the Virgo cluster and NGC 2784 group.
The origin of this difference is not yet understood, but might plausibly be driven by different formation or evolutionary processes in field and cluster environments.

In this work, we analyse the mass and surface brightness profiles of disc-type galaxies in the EAGLE simulations \citep{Schaye_et_al_15, Crain_et_al_15} in order to investigate the effect of environment on galaxy profile types, with the aim of understanding the suppressed fraction of Type~II galaxies in galaxy clusters.
In Section~\ref{sec:methods} we describe the EAGLE simulations and analysis of the simulations.
In Section~\ref{sec:profiles} we first discuss the origin of Type~II and III profiles in EAGLE galaxies, before exploring the the effect of environment on disc type frequencies in Section~\ref{sec:frequencies}.
Finally, we summarise the results in Section~\ref{sec:summary}.

\section{Methods} \label{sec:methods}

\subsection{EAGLE simulations} \label{sec:eagle}

The Evolution and Assembly of GaLaxies and their Environments (EAGLE) simulations are a suite of cosmological, hydrodynamical simulations of galaxy formation in the $\Lambda$ cold dark matter ($\Lambda$CDM) cosmogony \citep{Schaye_et_al_15, Crain_et_al_15}. 
The simulations were run with a highly modified version of the $N$-body Tree-PM smooth particle hydrodynamics (SPH) code \textsc{gadget3} \citep[last described by][]{Springel_05}, with updates to the SPH formulation, time-stepping and subgrid physics \citep[see appendix A of][]{Schaye_et_al_15}.
EAGLE includes subgrid routines describing radiative cooling \citep{Wiersma_Schaye_and_Smith_09}, star formation \citep{Schaye_and_Dalla_Vecchia_08}, stellar evolution and mass loss \citep{Wiersma_et_al_09}, the seeding and growth of black holes (BHs) via gas accretion and BH-BH mergers \citep{Rosas_Guevara_et_al_15}, and feedback associated with star formation \citep{Dalla_Vecchia_and_Schaye_12} and BH growth \citep{Booth_and_Schaye_09}.
The simulations adopt cosmological parameters consistent with a \citet{Planck_2014_paperXVI_short} cosmology, namely $\Omega_{\rm m} =  0.307$, $\Omega_\Lambda = 0.693$, $\Omega_{\rm b} = 0.04825$, $h = 0.6777$ and  $\sigma_8 =  0.8288$.
Dark matter structures are first identified by running the friends-of-friends (FOF) algorithm \citep{Davis_et_al_85} with a linking length 0.2 times the mean interparticle separation.
Bound galaxies (subhaloes) were identified within the FOF groups using the \textsc{subfind} algorithm \citep{Springel_et_al_01, Dolag_et_al_09}.

The stellar and BH feedback parameters are calibrated such that the simulations of cosmologically representative volumes reproduce the galaxy stellar mass function, galaxy sizes and BH masses at $z \approx 0$ \citep{Crain_et_al_15}.
Following this the EAGLE simulations reproduce a large range of observed galaxy population properties, making them ideal for comparisons with observed galaxy populations, including: the evolution of the galaxy stellar mass function \citep{Furlong_et_al_15} and sizes \citep{Furlong_et_al_17}, galaxy luminosities and colours \citep{Trayford_et_al_15}, galaxy morphologies \citep{Bignone_et_al_20}, cold gas properties \citep{Lagos_et_al_15, Lagos_et_al_16,Bahe_et_al_16,Marasco_et_al_16,Crain_et_al_17}, and the properties of circumgalactic and intergalactic absorption systems \citep{Rahmati_et_al_15,Rahmati_et_al_16,Oppenheimer_et_al_16,Turner_et_al_16,Turner_et_al_17}, while broadly reproducing the cosmic star formation rate density and relation between specific star formation rate and galaxy mass \citep{Furlong_et_al_15}.

This work makes use of both the EAGLE reference simulation (Ref-L100N1504) and the higher resolution `recalibrated' simulation (Recal-L025N0752; see \citealt{Schaye_et_al_15} for further details of the simulations).
EAGLE Ref-L100N1504 simulates a periodic volume with side length of 100 comoving Mpc using $1504^3$ gas and dark matter particles with initial baryonic masses $1.81 \times 10^6 \Msun$ and maximum gravitational softening length $0.7 \kpc$.
EAGLE Recal-L025N0752 simulates a periodic volume with a side length of 25 comoving Mpc using $752^3$ gas and dark matter particles with initial baryonic masses $2.26 \times 10^5 \Msun$ and maximum softening length $0.35 \kpc$.

\subsection{Sample selection and analysis} \label{sec:analysis}

We selected galaxies to have disc-type morphologies using the parameter $\kappaco$, the fraction of stellar kinetic energy invested in ordered co-rotation \citep{Correa_et_al_17, Thob_et_al_19}.
\citet{Correa_et_al_17} found that low $\kappaco$ galaxies ($\kappaco \sim 0.2$) tend to be spheroidal-shaped galaxies, while high $\kappaco$ galaxies ($\kappaco \sim 0.7$) tend to be disc-shaped galaxies.
Following \citet{Correa_and_Schaye_20}, we select disc-type galaxies to have $\kappaco > 0.35$.
With this selection, our sample ranges from S0s ($\kappaco \sim 0.4$) to late-type disc ($\kappaco \gtrsim 0.6$) galaxies \citep{Correa_et_al_19}.

In order to sample the radial profiles of the galaxies, we only consider well resolved galaxies with stellar masses $M_* > 10^{10} \Msun$, corresponding to galaxies with $>5\times 10^3$ stellar particles in the Ref-L100N1504 simulation and $>4 \times 10^4$ particles in the Recal-L025N0752 simulation.
We exclude galaxies in close proximity to significant neighbouring galaxies which result in major disturbances in their surface brightness profiles at the position of the neighbour galaxy, or obvious truncations resulting from the \textsc{subfind} algorithm and the lack of density contrast between the substructures.
This selection gives us a sample of 2290 galaxies for the reference simulation and 50 galaxies for the high resolution simulation.

For comparison with observed galaxy luminosity profiles \citep[i.e.][]{Erwin_et_al_12}, we convert stellar mass profiles of the simulated galaxies to SDSS $r$-band profiles using the \textsc{fsps} stellar population model \citep{Conroy_Gunn_and_White_09, Conroy_and_Gunn_10} assuming a \citet{Chabrier_03} initial stellar mass function and using the Miles spectral library \citep{Sanchez-Blazquez_et_al_06} and Padova isochrones \citep{Girardi_et_al_00, Marigo_and_Girardi_07, Marigo_et_al_08}.
We assume simple stellar populations for each stellar particle, and determine mass-to-light ratios for each particle by linearly interpolating from the grid in ages and total metallicities.
As a simple model for dust absorption, we assume stars with ages $<10 \Myr$ are fully embedded within an optically thick cloud \citep[e.g.][]{Charlot_and_Fall_00}.

Before classifying the $r$-band profiles we rotated all galaxies into face-on projections, such that they are comparable with face-on to moderately inclined samples of observed galaxies \citep[e.g.][]{Pohlen_and_Trujillo_06, Erwin_et_al_08, Gutierrez_et_al_11, Munoz-Mateos_et_al_13, Laine_et_al_14, Tang_et_al_20}.
To rotate the galaxies we calculate the spin vector for all stars between $2.5$ and $30 \kpc$ from the centre of potential each the galaxy.
The radius limits were chosen following \citet{Trayford_et_al_17}, such that the spin is dominated by the rotating disc component of the galaxy, if present, rather than bulge regions.

Radial surface brightness profiles for each galaxy were then created by using circular annuli about the centre of potential of the galaxy.
We use linearly increasing annuli, where the distance between inner and outer radii increases linearly with each annulus, beginning at $0.1\kpc$ and increasing in steps by $0.02 \kpc$, reaching a width of $1\kpc$ at a radius of $\approx 25\kpc$.
In this way the the densest, inner regions of each galaxy can be reasonably resolved, while retaining annuli in the outskirts of galaxies with areas large enough to avoid stochasticity due to particle resolution.

We classified the $r$-band surface brightness profile of each galaxy visually into exponential (Type~I), down- (Type~II) and up-breaking (Type~III), and composite profiles (e.g. Type~II break followed by a Type~III break, or the inverse), similar to the manor applied in observational works \citep[e.g.][]{Erwin_et_al_05, Pohlen_and_Trujillo_06, Erwin_et_al_08}.
We adopt an $r$-band surface brightness limit of $27 \Mag \Arcsec^{-2}$ for classifying the profiles, similar to observations in \citet{Erwin_et_al_12}.
In cases where multiple profile breaks are found (composite profiles) we adopt the break type at the smallest galactocentric radius, as an outer break might often fall below the surface brightness limit if a brighter limit was adopted (i.e. $< 27 \Mag \Arcsec^{-2}$).

\section{Origin of disc profile breaks in EAGLE galaxies} \label{sec:profiles}

In this section, we first investigate the origin of disc profile breaks with the Recal-L025N0752 simulation, which better samples the radial profiles due to its higher resolution.

\subsection{Type I: exponential disc} \label{sec:TypeI}

\begin{figure}
  \includegraphics[width=84mm]{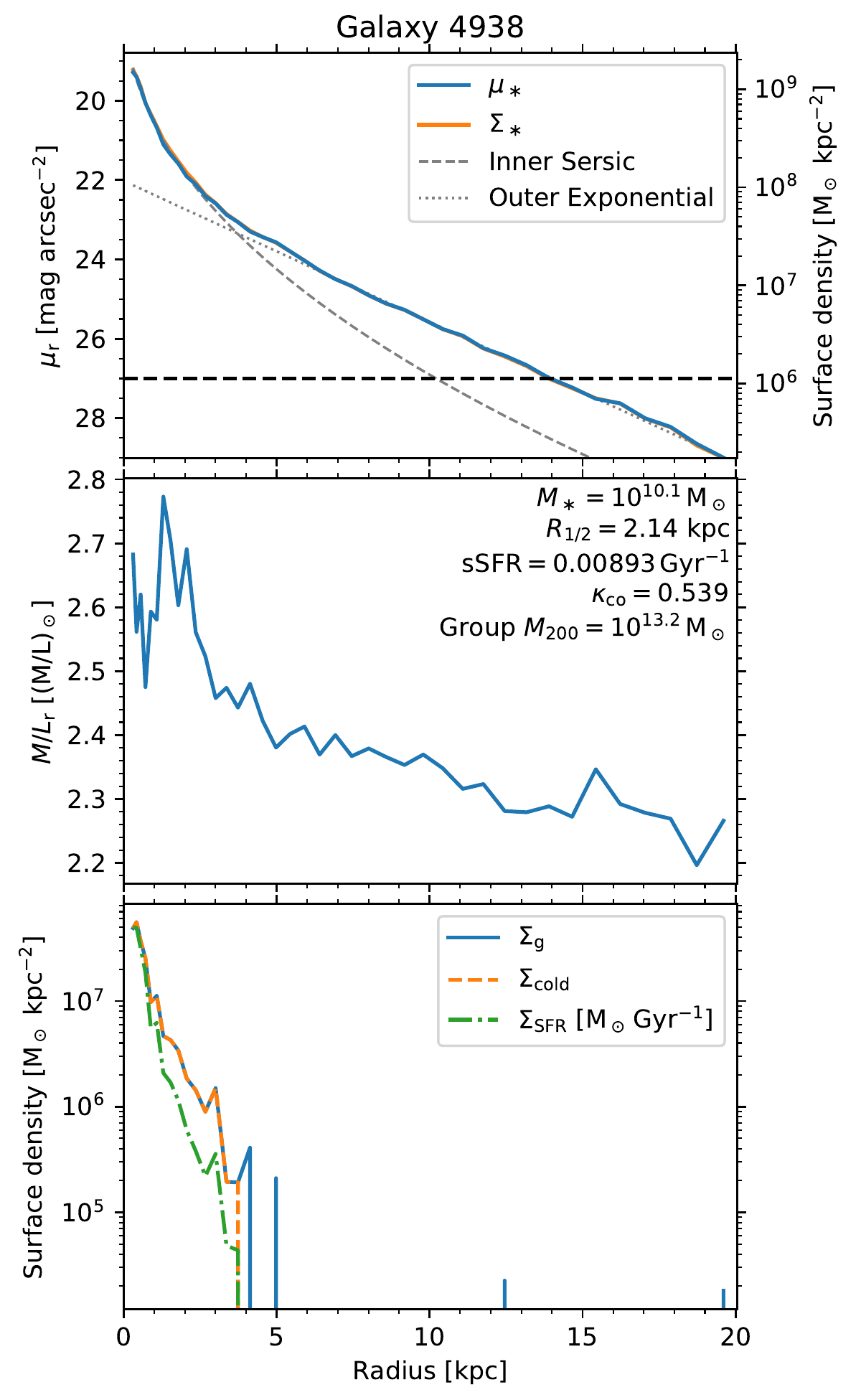}
  \includegraphics[width=84mm]{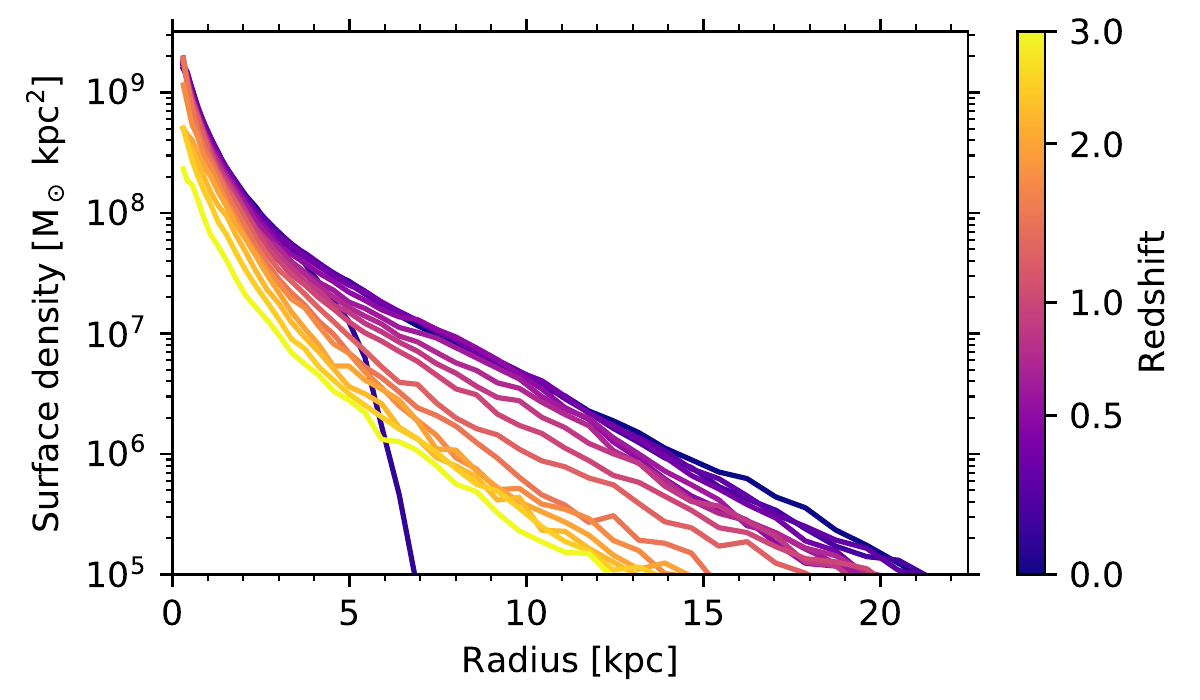}
  \caption{Example of a Type~I galaxy from the EAGLE Recal-L025N0752 simulation (GalaxyID$=$4938).  The top panel shows the projected $r$-band surface brightness (solid blue line, left axis) and mass density profiles (solid orange line, right axis).  The horizontal dashed line shows the adopted surface brightness limit of $27 \Mag \Arcsec^{-2}$.  The grey dashed line shows a \citet{Sersic_63} profile fit to the inner $2.5 \kpc$ of the surface brightness profile, while the grey dotted line shows an exponential profile fit to the outer quarter of profile (up to $27 \Mag \Arcsec^{-2}$).  The second panel shows the radial profile of the $r$-band mass-to-light ratio.  Text in the second panel displays information about the galaxy (i.e. stellar mass, half-mass radius, specific star formation rate, kinematic morphology ($\kappaco$) and group $M_{200}$).  The third panel shows the surface density profiles for the total gas mass ($\Sigma_\mathrm{g}$), star-forming (cold) gas mass ($\Sigma_\mathrm{cold}$) and star formation rate ($\Sigma_\mathrm{SFR}$, in $\Msun \Gyr^{-1} \kpc^{-2}$).  The bottom panel shows the evolution of the mass surface density profile for the galaxy from $z=3$ to $z=0$.}
  \label{fig:TypeI}
\end{figure}

As a reference to contrast with the truncated (Section~\ref{sec:TypeII}) and anti-truncated (Section~\ref{sec:TypeIII}) galaxy profiles, we first show an example of a Type~I galaxy from the EAGLE Recal-L025N0752 simulation in Fig.~\ref{fig:TypeI}.
The surface brightness profile shows a clear inner bulge (well fit by a \citealt{Sersic_63} profile), as well as an outer exponential disc.
The galaxy shows very similar light and mass profiles (upper panel), due to its very flat mass-to-light ratio profile (second panel). 

The evolution of the density profile for the galaxy in is shown in the bottom panel of Fig.~\ref{fig:TypeI}.
The bulge of the galaxy forms by $z \sim 1.5$, while the disc continues to form to $z \approx 0.2$.
The evolution of the density profile for this galaxy is relatively smooth due to the absence of any significant galaxy accretion events.
The truncated profile at $z=0.1$ occurs where the \textsc{subfind} algorithm assigns the outer particles of this galaxy to a more massive galaxy during a close passage in a galaxy group (see \citealt{Knebe_et_al_11} and \citealt{Muldrew_et_al_11} for further discussion on the difficulties of subhalo detection in a high background density). These outer particles are reassigned to the galaxy in the $z=0$ snapshot.
As a result of the interaction, the galaxy has been stripped of most of its gas beyond $5 \kpc$ at $z=0$ (third panel).

We note that not all Type~I galaxies follow identical formation histories. We find examples of Type~I galaxies that form outside-in, as well as those that retain a Type~I profile despite significant late galaxy mergers.

\subsection{Type II: truncated disc} \label{sec:TypeII}

\begin{figure*}
  \includegraphics[width=84mm]{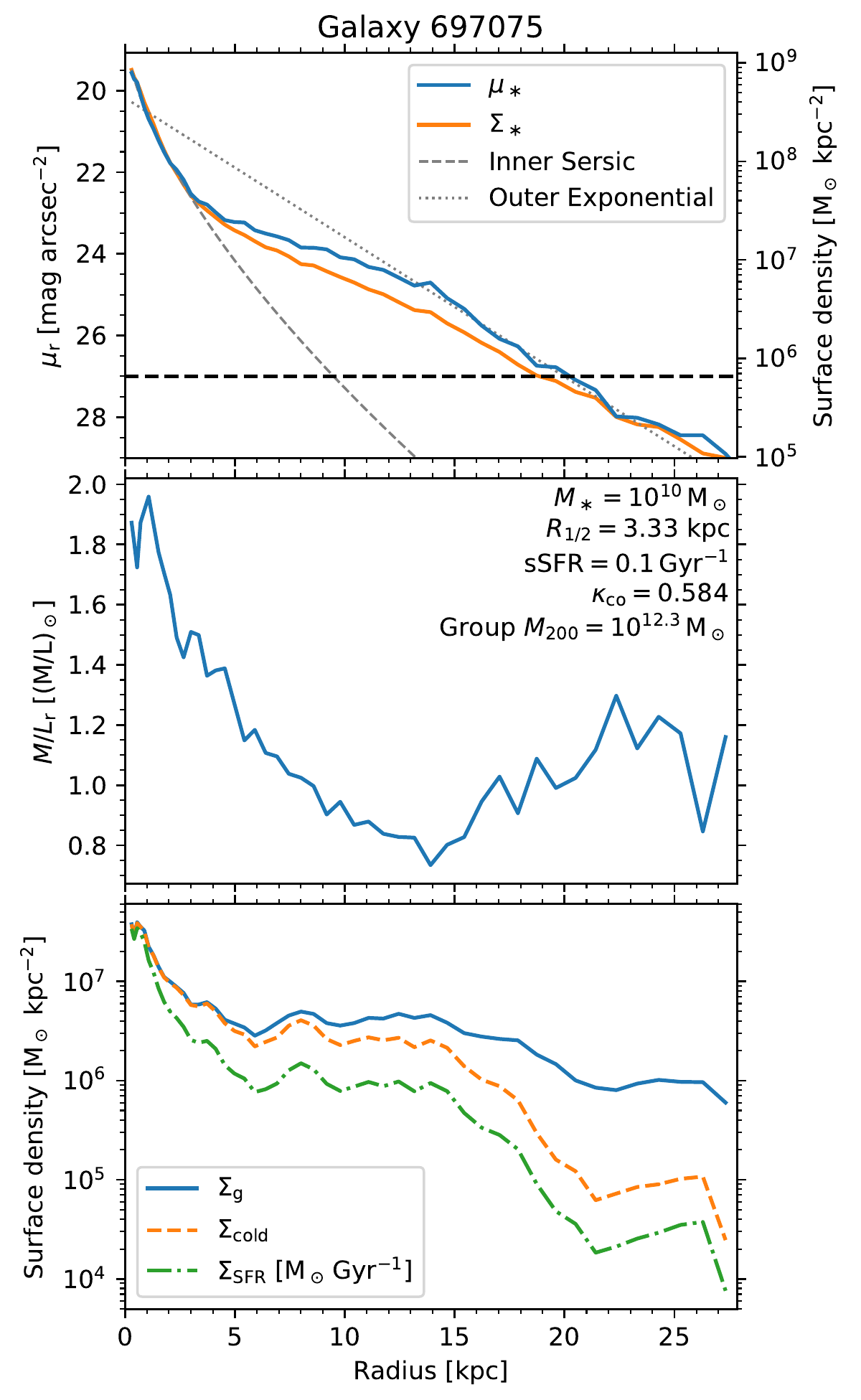}
  \includegraphics[width=84mm]{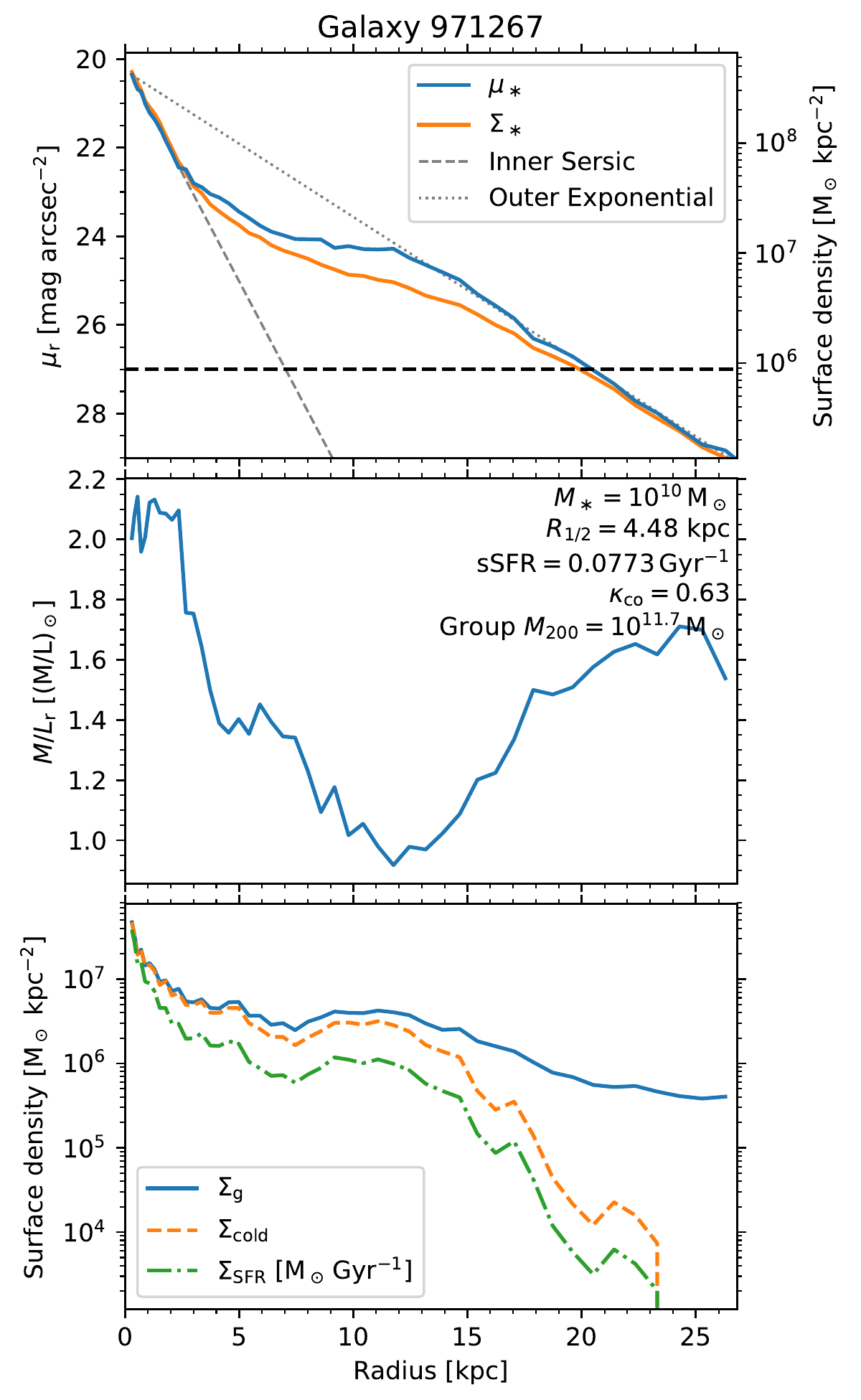}
  \includegraphics[width=84mm]{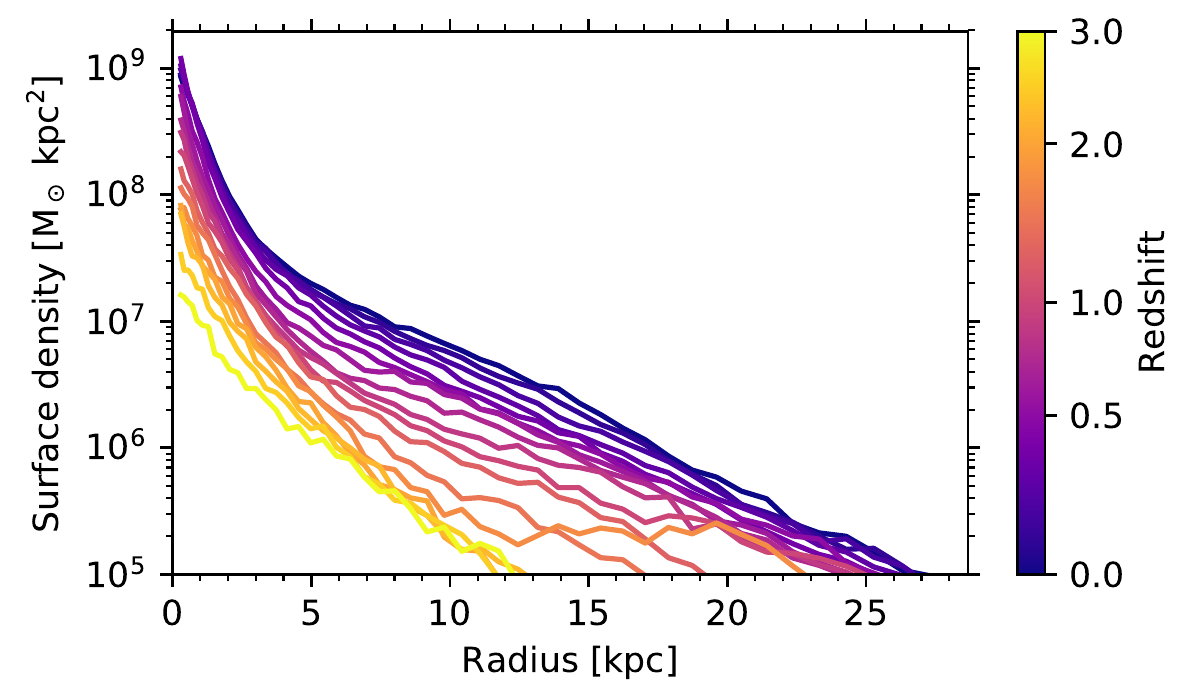}
  \includegraphics[width=84mm]{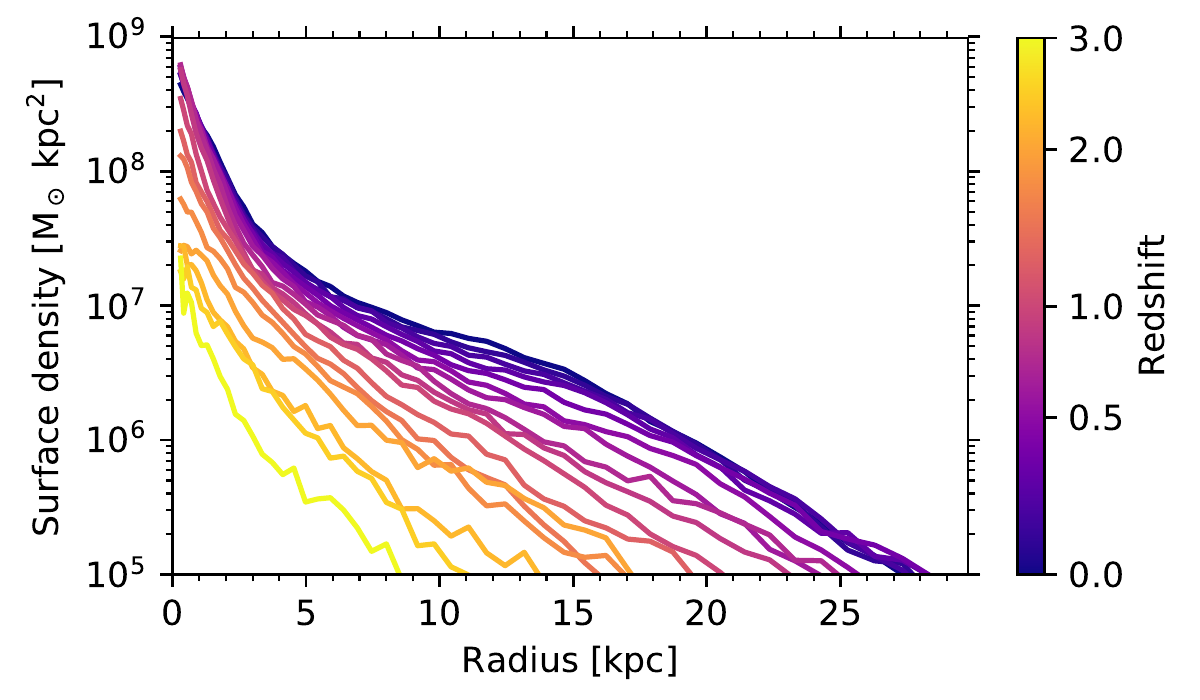}
  \caption{Examples of two Type~II galaxies from EAGLE Recal-L025N0752. Panels in the figure are as in Fig.~\ref{fig:TypeI}. For galaxy 697075 (left panels) the profile truncation mainly exists in the luminosity profile. For galaxy 971267 (right panels) the profile truncation occurs in both the mass and luminosity profiles.}
  \label{fig:TypeII}
\end{figure*}

We show two examples of EAGLE galaxies with truncated disc profiles in Figures \ref{fig:TypeII} (though the results apply more generally to all Type~II galaxies found in the simulation).
Based on their kinematic morphology ($\kappaco$), both examples would be considered spiral galaxies.
For both galaxies, the Type~II break in the disc profiles ($R_\mathrm{break} \approx 14 \kpc$ for Galaxy 697075, $R_\mathrm{break} \approx 12 \kpc$ for Galaxy 971267) occurs near the minimum in the mass-to-light ratio profiles (both galaxies show the characteristic `U' shaped $M/L$ profiles).
In both cases, the low mass-to-light ratios are due to the younger stellar populations near the disc break (median stellar particle ages $\sim 2 \Gyr$ younger than the galaxy as a whole).
Beyond the break radii, the mass-to-light ratios increase due to the transition to older, more metal-poor stellar populations (i.e. the stellar halo).

In both galaxies, the minimum in the $M/L$ ratio profile is driven by a rapid drop-off in the cold gas mass and SFR beyond the disc break radius (third panels from the top).
The cold gas mass and SFR density show a drop-off once the total gas surface density falls below $\sim 10^{6.5} \Msun \kpc^{-2}$, which result from the density threshold for star formation used in the EAGLE star formation model \citep{Schaye_04, Schaye_and_Dalla_Vecchia_08}. 
Beyond the drop-off in the cold gas density/SFR, the mass-to-light ratio of the stellar population rises, thus creating the appearance of a `break' in the disc profile.

In the case of Galaxy 697075 (left panels in Fig.~\ref{fig:TypeII}), the broken disc profile occurs in surface brightness profile only; in the mass density profile the galaxy has a Type~I (exponential) disc profile (similar to the galaxy in Fig.~\ref{fig:TypeI}).
Thus for this galaxy, the Type~II surface brightness profile is a result of the stellar population gradient, not the mass profile itself. A similar result was found in the simulation by \citet{Sanchez-Blazquez_et_al_09}.

In the case of Galaxy 971267 (right panels in Fig.~\ref{fig:TypeII}), it also features a Type~II break in its mass density profile as well as its brightness profile.
The evolution of the surface density profile for this galaxy is shown in the bottom right panel of Fig.~\ref{fig:TypeII}.
This galaxy has only very minor mergers (stellar mass merger ratios less than 1:10) since $z \sim 2$, and thus the density profile evolution is driven purely by in-situ star formation.
The development of a Type~II mass profile for this galaxy occurs relatively late in its evolution, from $z \lesssim 0.5$.
The density profile for the very outer disc ($\gtrsim 18 \kpc$) has remained nearly constant during this time, with the break radius progressively decreasing as the inner disc ($\sim 5$-$15 \kpc$) continues to grow.
However, Galaxy 971267 appears to be an exception, as most Type~II galaxies in the Recal-L025N0752 simulation have near exponential mass profiles.
Where a Type~II break does occur in the mass profile, it is less pronounced than the break in the light profile (as seen for Galaxy 971267 in Fig.~\ref{fig:TypeII}).

These results on the origin of Type~II disc profiles are in good agreement with that from observed Type~II disc galaxies, which also have break radii that coincide with minima in colour, age and $M/L$ \citep{Azzollini_et_al_08a, Bakos_et_al_08, Zheng_et_al_15, Ruiz-Lara_et_al_16}.
Similarly, using $M/L$ profiles to convert surface brightness to mass profiles shows that Type~II breaks are far weaker, if present at all, in the mass profiles of observed galaxies \citep{Bakos_et_al_08, Tang_et_al_20}.
The results from the EAGLE simulations also agree with that of previous simulations, showing that Type~II disc truncations result from a drop-off in the cold gas density near the break radius in star-forming galaxies \citep{Roskar_et_al_08, Martinez-Serrano_et_al_09, Sanchez-Blazquez_et_al_09}.

We note that for barred galaxies the Type~II breaks are often coincident with an outer ring structure in the galaxy, likely related to the outer Lindblad resonance of the bars \citep{Pohlen_and_Trujillo_06, Erwin_et_al_08, Munoz-Mateos_et_al_13}.
The Type~II galaxies from the Recal-L025N0752 volume did not show obvious bars or rings, though the galaxy sample size is also rather small (50 galaxies). 
Extending this analysis to investigate the connection between bars and Type~II breaks in simulated galaxies would be a worthwhile avenue for future work.

\subsection{Type III: anti-truncated disc} \label{sec:TypeIII}

\begin{figure}
  \includegraphics[width=84mm]{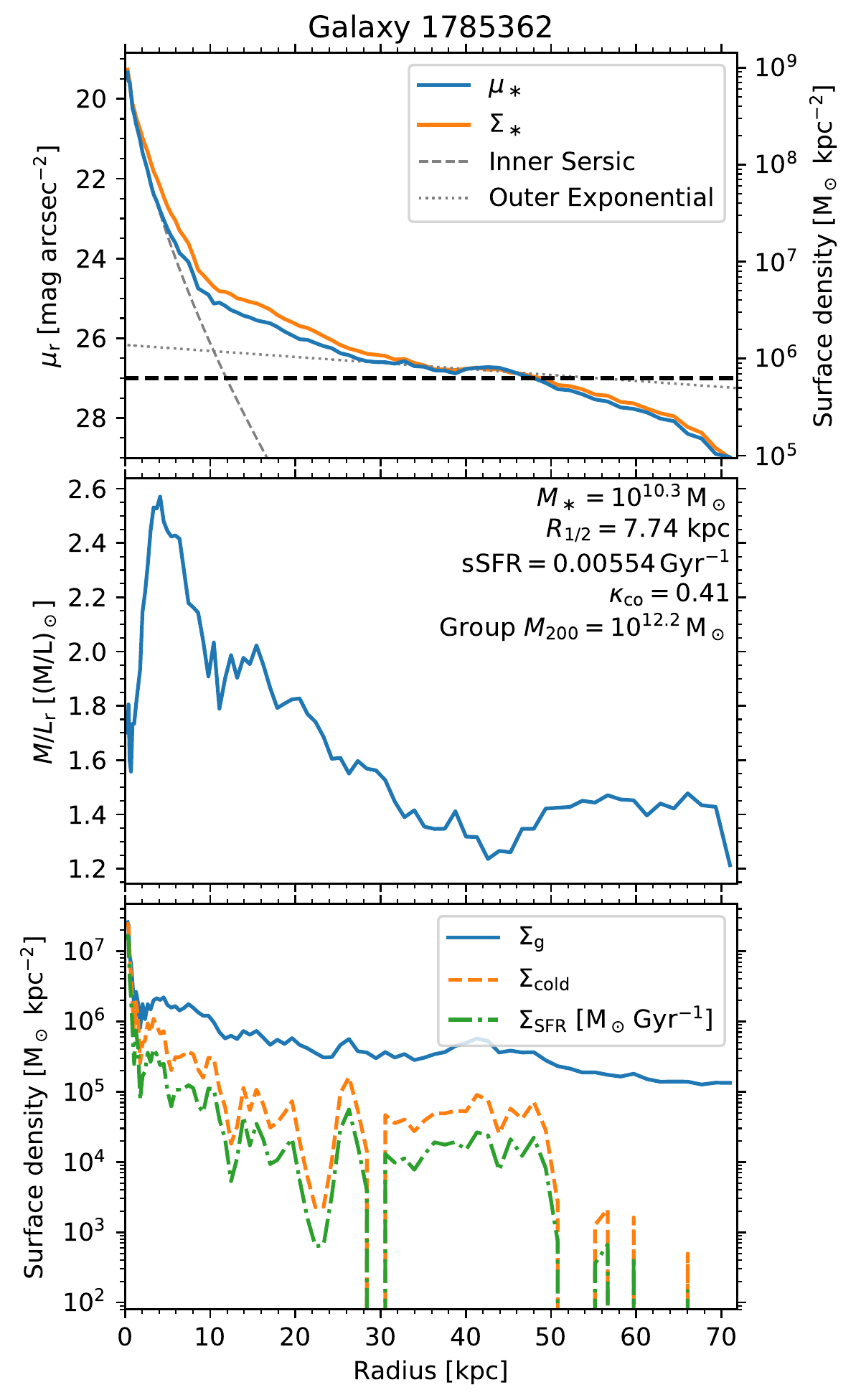}
  \includegraphics[width=84mm]{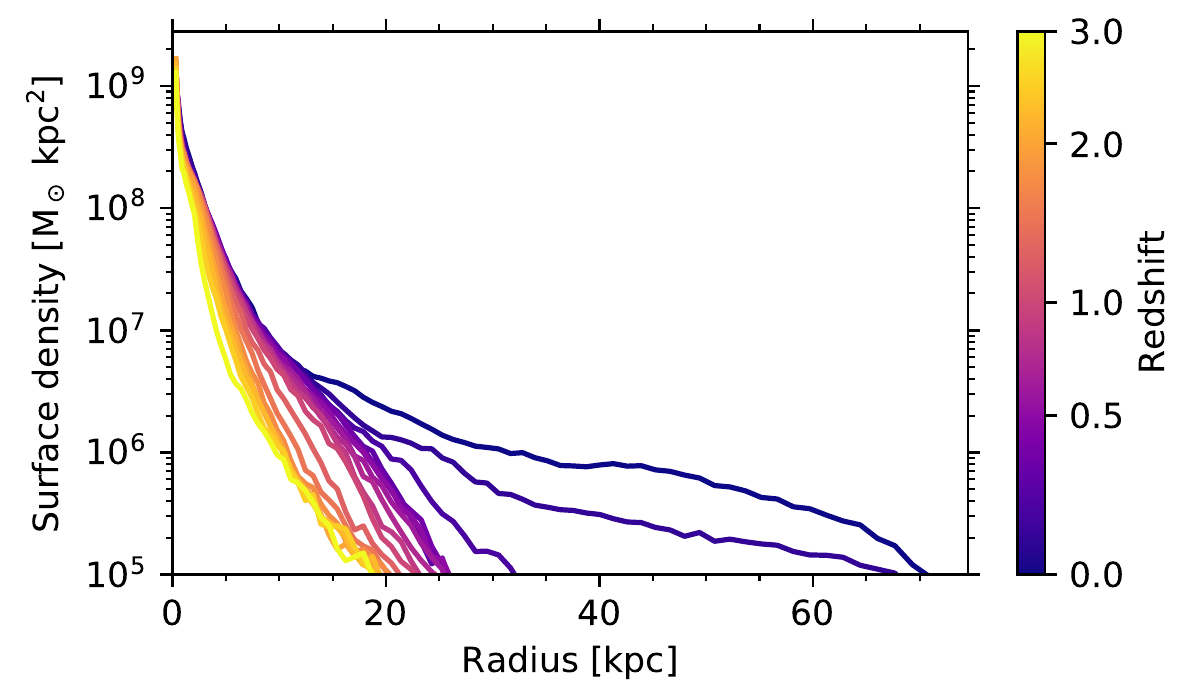}
  \caption{Example of a Type~III (anti-truncated) galaxy formed from a late ($z \approx 0.1$) minor merger (stellar masses $\approx 2 \times 10^{10}$ and $\approx 3 \times 10^{9} \Msun$). Panels in the figure are as in Fig.~\ref{fig:TypeI}.}
  \label{fig:TypeIII_merger}
\end{figure}

\begin{figure*}
  \includegraphics[width=84mm]{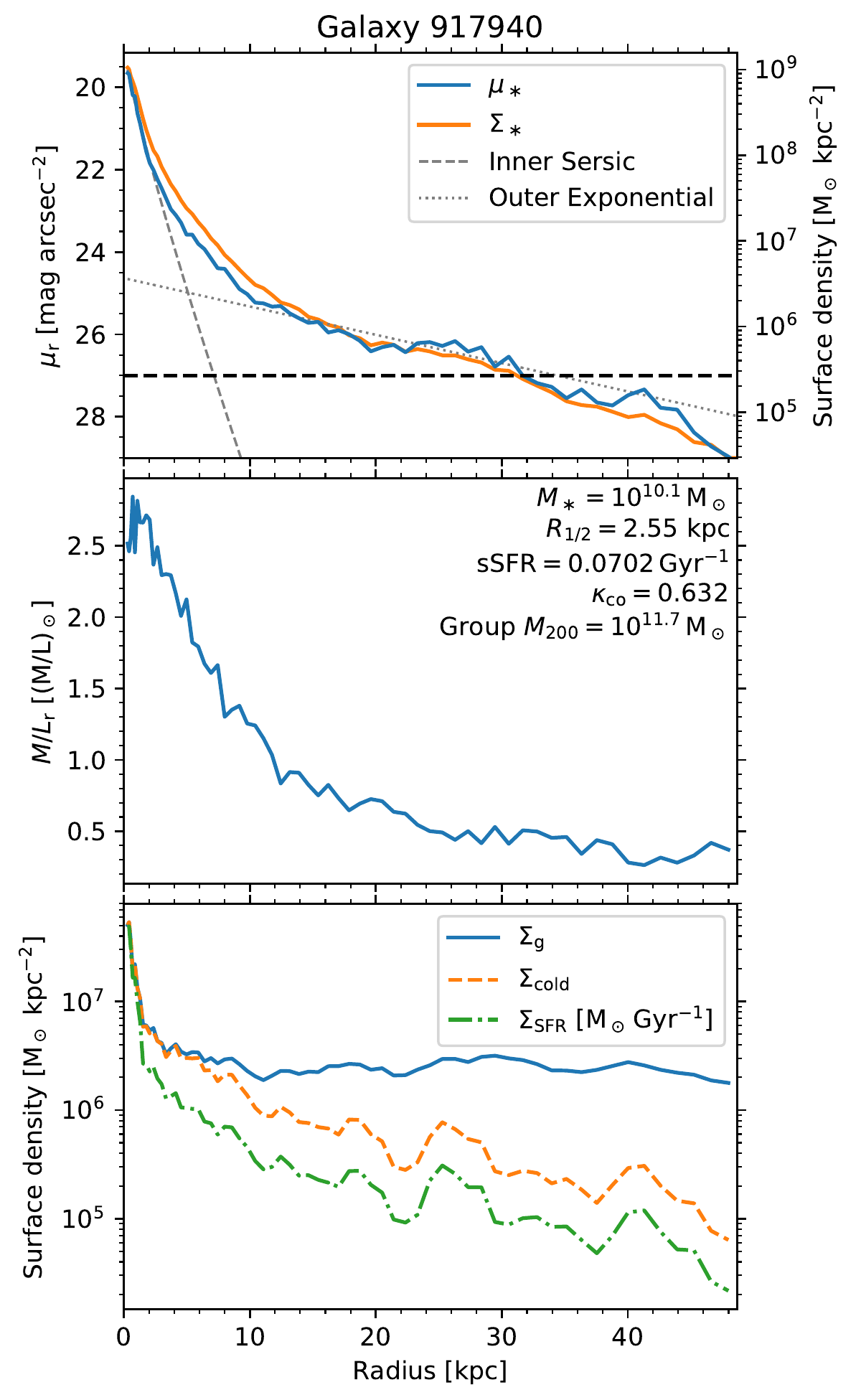}
  \includegraphics[width=84mm]{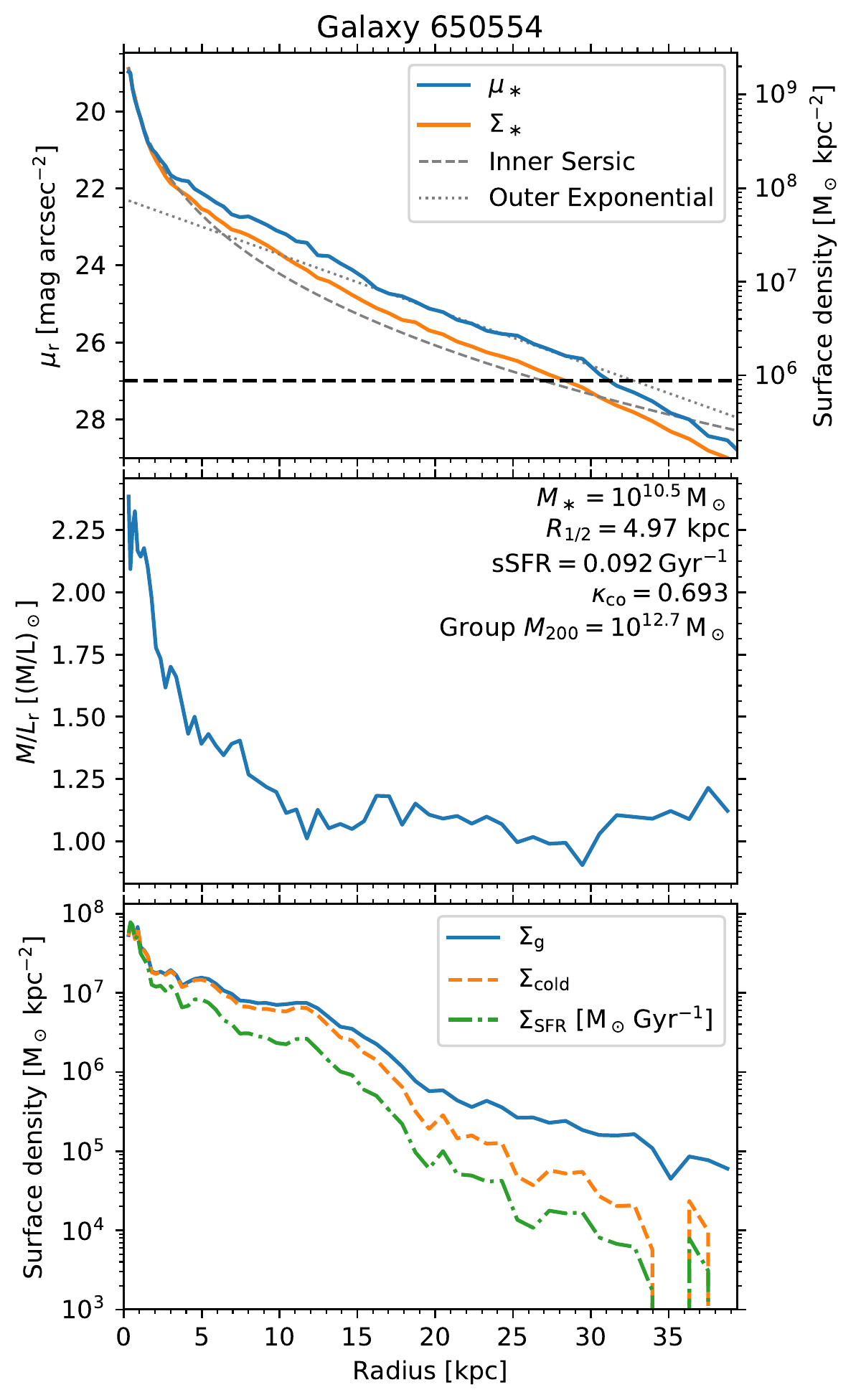}
  \includegraphics[width=84mm]{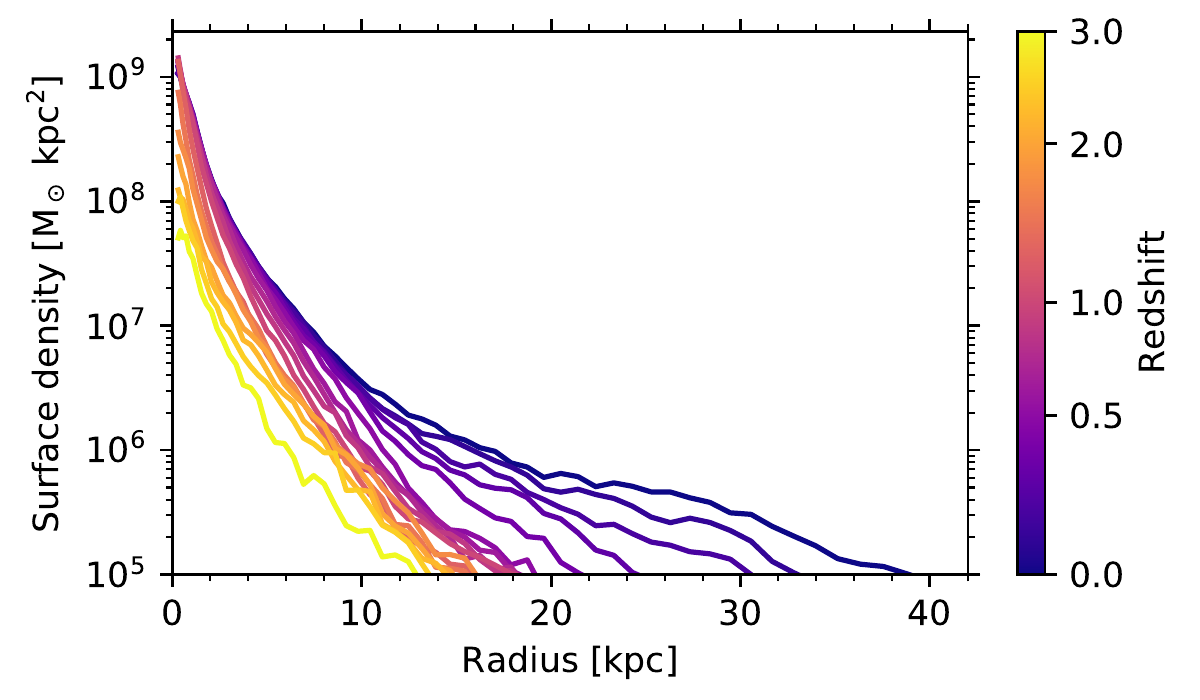}
  \includegraphics[width=84mm]{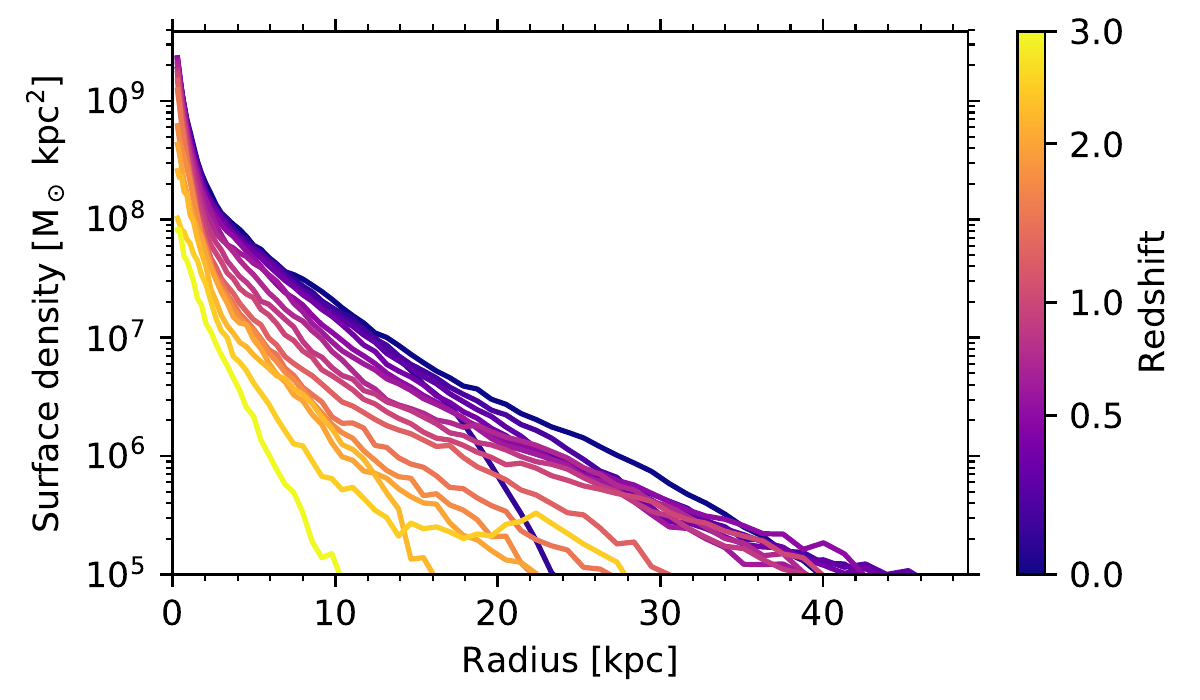}
  \caption{Two examples of Type~III profiles from EAGLE Recal-L025N0752 formed via star-forming discs. Panels in the figure are as in Fig.~\ref{fig:TypeI}. For galaxy 917940 (left panels) the outer disc profile ($\gtrsim 10 \kpc$) builds up late due to an extended star-forming disc. For galaxy 650554 (right panels) the outer disc profile ($15$-$30 \kpc$) forms prior to the inner disc ($5$-$15 \kpc$).}
  \label{fig:TypeIII_SF}
\end{figure*}

While Type~II galaxies in EAGLE largely do not show disc profile breaks in their mass profiles, the case is different for Type~III galaxies. We find such anti-truncated discs generally show clear breaks in both their mass and brightness profiles, and thus cannot simply be explained by radial gradients in their stellar populations.
Additionally, while Type~II galaxies in the EAGLE simulations appear to have a single origin relating to drop-offs in their cold gas/SFR profiles, the origin of Type~III profiles is more varied.

For some galaxies, the anti-truncation is a clear result of stars accreted during a galaxy merger \citep[e.g.][]{Bekki_98, Bournaud_et_al_05}, whether a minor or major merger. 
We show one such example in Fig.~\ref{fig:TypeIII_merger}. 
This galaxy undergoes a minor merger with a satellite galaxy with $\sim 15$ per cent of its stellar mass at $z \approx 0.1$.
Prior to this merger, the galaxy appeared simply to have a Type~I (exponential) profile.
We do not find any clear indicators in the $M/L$ or age profiles of merger-created Type~III galaxies, since it depends upon the relative properties of galaxies involved in the merger and any subsequent star formation.
For this example, $M/L$ decreases gradually with increasing radius after peaking at $4 \kpc$.
As we will see in Section~\ref{sec:frequencies} (Fig.~\ref{fig:types_Mstar}), Type~III profiles are very common for very massive galaxies with $M_\ast > 10^{11} \Msun$, which build a significant amount of their stellar mass via mergers \citep{Rodriguez-Gomez_et_al_16, Qu_et_al_17, Clauwens_et_al_18, Tacchella_et_al_19, Davison_et_al_20}.
Anti-truncated profiles formed by mergers may account for the Type~III-s profiles (disc embedded within an outer spheroid) observed from some galaxies \citep{Erwin_et_al_05, Erwin_et_al_08}.

Alternatively, the anti-truncation can be due to an extended star-forming disc at late cosmological times. 
This mechanism is similar to that found by \citet{Wang_et_al_18}, and related to the up-bending SFR profiles found by \citet{Tang_et_al_20}.
In the left panels of Fig.~\ref{fig:TypeIII_SF} we show one example, where the anti-truncation is caused by a nearly flat star formation profile at $>10 \kpc$.
For this galaxy the inner, steeper profile (radii from 3-10 kpc), is largely in place by $z \sim 0.5$, while the outer, flatter disc profile grows from $z=0.5$ to $z=0$.
Interestingly, this outer disc then features a Type~II break at $\approx42 \kpc$, at which point the cold gas surface density begins to drop, but the break is fainter than the adopted surface brightness limit of $27 \Mag \Arcsec^{-2}$.

In other cases the inverse is true, the inner disc builds up at a later time than the outer disc.
We show one case in the right panels of Fig.~\ref{fig:TypeIII_SF}.
In this galaxy, the outer disc ($\gtrsim 15 \kpc$) is nearly in place by $z \sim 1$, while the inner disc continues to grow until the present time. 
This case is reminiscent of Type~II galaxies, in that the star-forming gas disc drops off significantly beyond $15 \kpc$, but without the corresponding truncation in the brightness profile.
Anti-truncated profiles formed by star formation would be expected to correspond to the subset of profiles where the break is a continuation of the disc \citep[Type~III-d, ][]{Erwin_et_al_05, Erwin_et_al_08}.

\citet{Roediger_et_al_12} and \citet{Wang_et_al_18} have also suggested that galaxy harassment in galaxy clusters may play a role in the formation of Type~III profiles. However, due to the infrequency of snapshots for the simulations (29 between $z=20$ and $z=0$), we cannot directly test such an origin.
Finally, we note that some processes for Type~III profile formation are not exclusive, and thus in some cases more than one effect may be at play within the one galaxy (e.g. accretion of both stars and gas during a merger, where the accreted gas subsequently fuels further star formation in the disc).

\section{Effect of environment on disc profile types} \label{sec:frequencies}

In this section, we investigate the light profiles of galaxies in the Ref-L100N1504 simulation, which provides better sampling of galaxies in different environments due to the larger simulation volume (at the expense of poorer sampling of the radial profiles due to the lower resolution).
Uncertainties on the frequencies in this section were calculated using binomial statistics and show $\pm 1 \sigma$ uncertainties.

\subsection{Correlations with mass, morphology and environment} \label{sec:mass_morphology}

\begin{figure}
  \includegraphics[width=84mm]{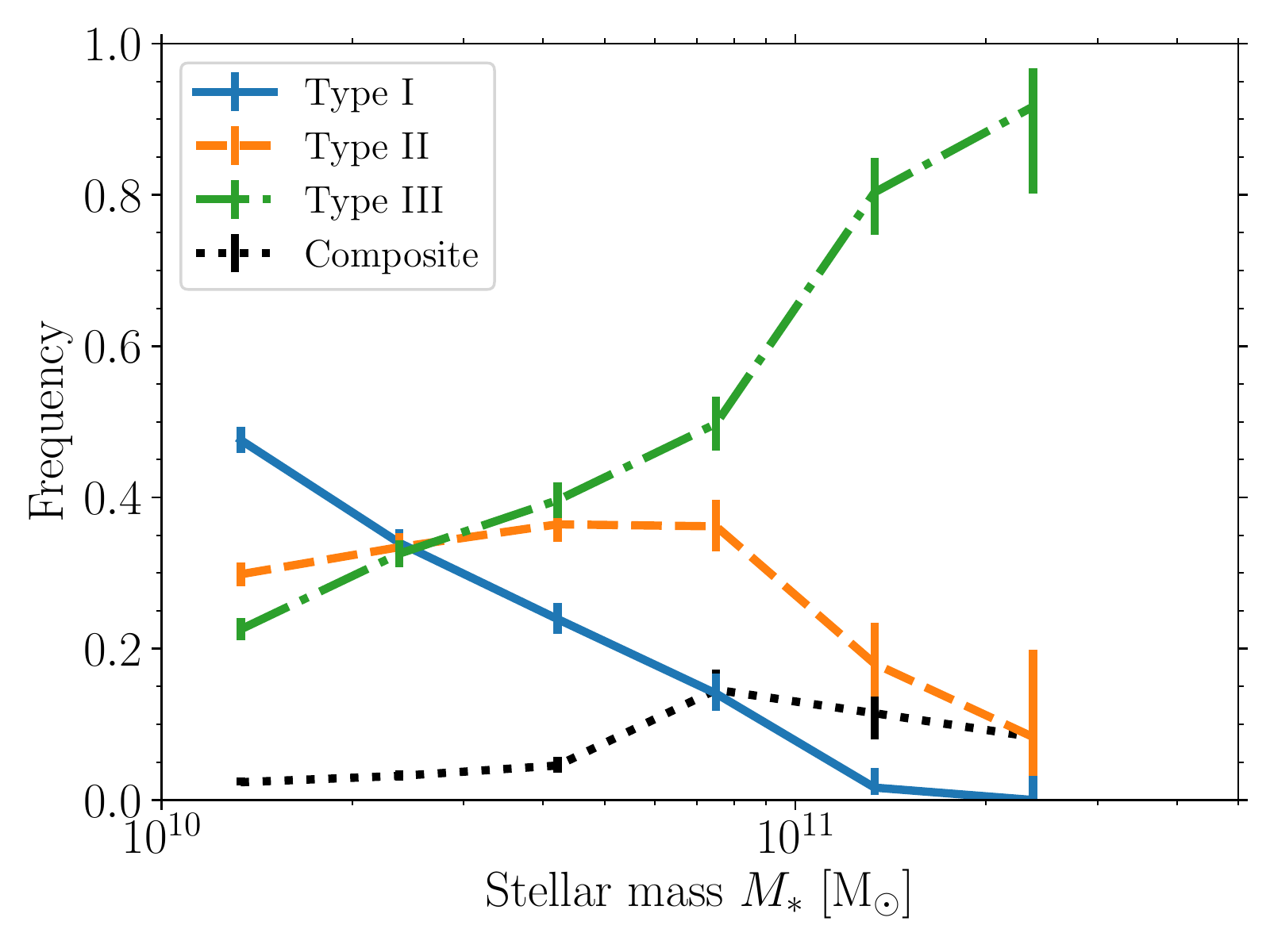}
  \caption{The frequency of galaxy profile types as a function of galaxy stellar mass in the EAGLE Ref-L100N1504 simulation. The fraction of Type~I (exponential) profiles decreases with increasing stellar mass, while the fraction of and Type~III (anti-truncated) profiles increases with stellar mass, in agreement with observed galaxies \citep{Laine_et_al_16, Tang_et_al_20}. The black dotted line shows the fraction of composite type profiles (Type~II+III or III+II). Errorbars show the $1\sigma$ uncertainties from binomial statistics.}
  \label{fig:types_Mstar}
\end{figure}

\begin{figure}
  \includegraphics[width=84mm]{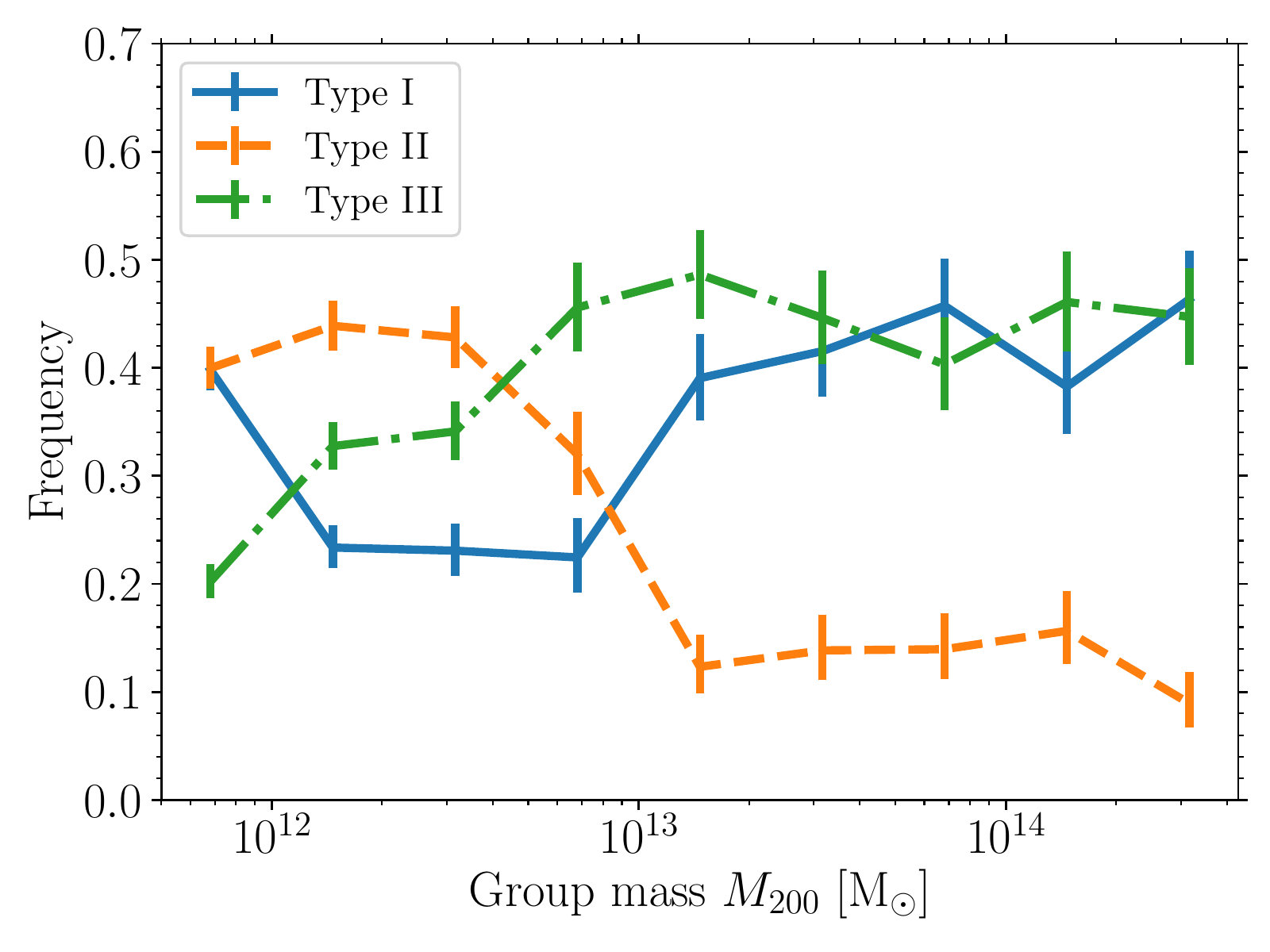}
  \caption{The frequency of galaxy disc types as a function of group mass ($M_{200}$) in the EAGLE Ref-L100N1504 simulation. The fraction of Type~II galaxies is highest in `field' galaxies ($M_{200} < 10^{13} \Msun$), then drops of significantly in groups with $M_{200} > 10^{13} \Msun$. Errorbars show the $1\sigma$ uncertainties from binomial statistics.}
  \label{fig:types_M200}
\end{figure}

\begin{figure*}
  \includegraphics[width=\textwidth]{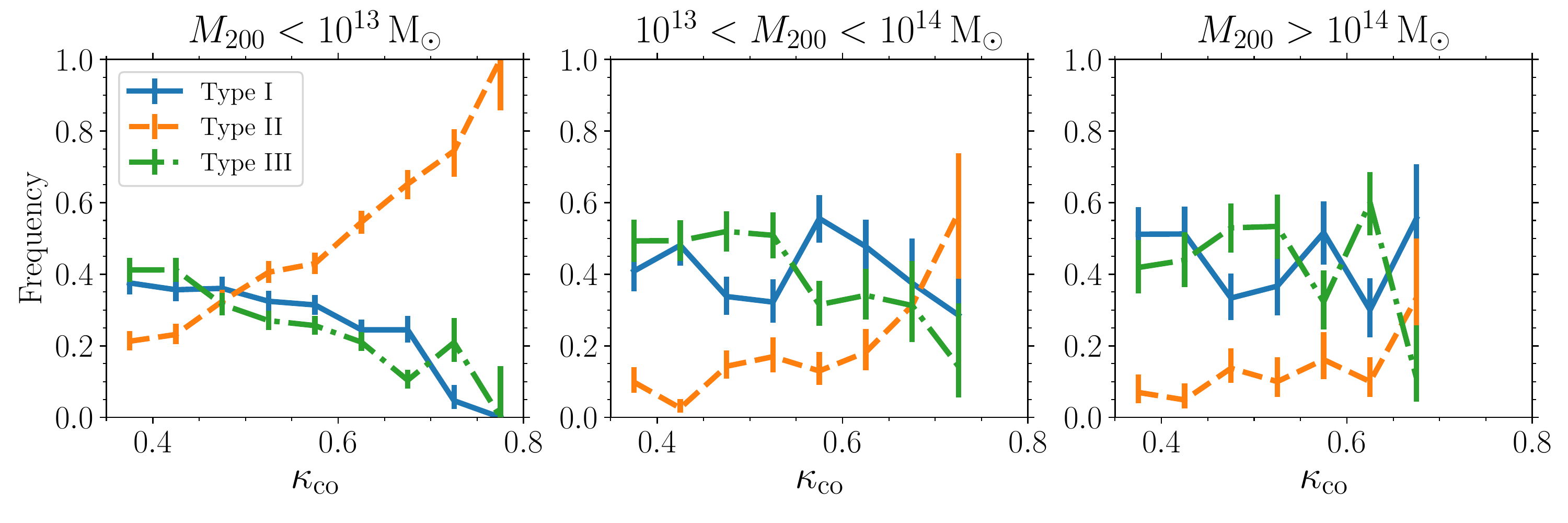}
  \caption{Frequency of galaxy disc types as a function of kinematic morphology, $\kappaco$ (fraction of stellar kinetic energy invested in ordered co-rotation). In this classification, S0 galaxies have $\kappaco \sim 0.4$ and Sbc or later corresponds to $\kappaco \gtrsim 0.6$. The galaxy sample is split into three different environments: `field galaxies' ($M_{200} < 10^{13} \Msun$), `galaxy groups' ($10^{13} < M_{200}/\Msun < 10^{14}$) and `galaxy clusters' ($M_{200} > 10^{14} \Msun$). Errorbars show the $1\sigma$ uncertainties from binomial statistics.}
  \label{fig:types_Kappa}
\end{figure*}

In Fig.~\ref{fig:types_Mstar}, we first show the relationship between disc type and galaxy stellar mass for the whole sample of simulated galaxies. 
Though the exact fractions differ (between both observational studies as well as the simulations), the results are in qualitative agreement with the trends found by \citet{Laine_et_al_16} and \citet{Tang_et_al_20} showing that the fraction of Type~III galaxies increases with galaxy mass, while the fraction of Type~I galaxies decreases with increasing galaxy mass.
The fraction of Type~III galaxies in the simulations shows a steeper trend with mass than the observations.
For masses $10^{10.5}$-$10^{11} \Msun$ the simulations are in good agreement with the \citet{Tang_et_al_20} results (Type~III fraction $\sim45$ per cent), but for masses $10^{10}$-$10^{10.5} \Msun$ are in better agreement with the \citet{Laine_et_al_16} results (Type~III fraction $\sim25$ per cent).
As discussed in Section~\ref{sec:TypeIII}, the increase of Type~III galaxies with galaxy mass can be understood due to the increasing contribution of mergers to galaxy growth, either through the direct accretion of stars creating a profile break, or via the accretion of gas to form extended star-forming discs.

The fraction of Type~II galaxies is relatively constant for stellar masses $<10^{11} \Msun$, at around 30-35 per cent, but drops off significantly at higher masses.
This fraction is slightly lower than observed \citep[$\approx$40 per cent,][]{Laine_et_al_16, Tang_et_al_20}, but is similarly independent of galaxy mass within the mass range $10^{10}$-$10^{11} \Msun$.
The lower fraction of Type~II galaxies in the EAGLE simulations relative to observed galaxies, and thus the slightly higher fraction of Type~I galaxies, might potentially be due to the lower than observed star formation rates of EAGLE galaxies \citep{Furlong_et_al_15}.
However, we note that the fraction of Type~II profiles in field galaxies ($M_{200} \sim 10^{12} \Msun$, see Fig.~\ref{fig:types_M200}) is similar to the observed fractions ($\sim$40 per cent).

We find that $4.4^{+0.5}_{-0.4}$ per cent of the simulated galaxies have composite profiles (i.e. Type~II+III or III+II), somewhat lower than the $8 \pm 2$ per cent found by \citet{Gutierrez_et_al_11}.
However the fraction of composite types is dependent on galaxy stellar mass, increasing from $3.1 \pm 0.4$ per cent for $M_\ast < 10^{10.75} \Msun$ to $13 \pm 2$ per cent for $M_\ast > 10^{10.75} \Msun$.
Therefore, a galaxy sample weighted to higher masses would obtain a higher fraction of composite profiles.
These composite profiles have a number of origins.
For Type~II+III galaxies, the inner Type~II break can be due to a compact, truncated star-forming disc (as in Section~\ref{sec:TypeII}) or a bar \citep[c.f.][]{Erwin_et_al_08}.
Type~III+II galaxies can result from very extended star-forming disc, bearing a relation to Type~III galaxies formed via a similar manner (see Section~\ref{sec:TypeIII} and Fig.~\ref{fig:TypeIII_SF}).
Of the galaxies with composite types, $73^{+4}_{-5}$ per cent are Type~II+III.

In Fig.~\ref{fig:types_M200} we compare the relationship between disc profile type and group mass ($M_{200}$ of the FOF group in which the galaxy resides at $z=0$).
The fraction of Type~II galaxies shows a clear decrease in denser environments (increasing $M_{200}$), consistent with the findings of \citet{Erwin_et_al_12} and \citet{Roediger_et_al_12}.
In contrast, the fraction of Type~III galaxies increases with group mass (from $10^{12}$ to $10^{13} \Msun$), while the fraction of Type~I galaxies is lowest at $M_{200} \sim 10^{12.5} \Msun$ and otherwise remaining at $\approx$40 per cent at lower and higher group masses.
We will further discuss the origin of these trends in Section~\ref{sec:TypeII_in_clusters}.

Next, we investigate the trend of disc profile type with morphology (via the kinematic indicator $\kappaco$) and environment in Fig.~\ref{fig:types_Kappa}.
\citet{Correa_et_al_19} found that $\kappaco$ correlates reasonably well with visual morphology (see their section 2.3). In this classification, S0 galaxies have $\kappaco \sim 0.4$ while Sbc or later corresponds to $\kappaco \gtrsim 0.6$.
We divide the sample by halo mass ($M_{200}$) of the FOF group into `field galaxies' ($M_{200} < 10^{13} \Msun$), `galaxy groups' ($10^{13} < M_{200}/\Msun < 10^{14}$) and `galaxy clusters' ($M_{200} > 10^{14} \Msun$).
The galaxy cluster sample has similar halo masses to the Virgo cluster \citep[$M_{200} \approx 4 \times 10^{14} \Msun$,][]{McLaughlin_99}.

The trend of disc type with morphology for `field' galaxies is in good agreement with observations of field galaxies \citep[e.g.][]{Gutierrez_et_al_11, Tang_et_al_20}: the frequency of Type~II galaxies increases as galaxies become more disc dominated, while the fraction of Type~I and III galaxies decreases with increasing $\kappaco$.

For S0-like galaxies with $0.35 < \kappaco < 0.45$, the fraction of Type~II galaxies decreases from $22 \pm 2$ per cent in field galaxies, to $6^{+3}_{-2}$ per cent in galaxy clusters.
Similarly, for very disc dominated galaxies with $\kappaco > 0.6$ the fraction of Type~II galaxies decreases from $61 \pm 2$ per cent in field galaxies, to $21^{+7}_{-6}$ per cent in galaxy clusters.
These results are in good agreement with the observations of \citet{Erwin_et_al_12} and \citet{Roediger_et_al_12}, who found a lack of Type~II S0 galaxies and a suppressed fraction of Type~II disc galaxies in the Virgo cluster, respectively.
In the Fornax cluster, a similarly suppressed fraction of Type~II disc galaxies (38 per cent) was found by \citet{Raj_et_al_19}.
A number of Fornax cluster S0 galaxies were found to have Type~II breaks, largely associated with inner bar structures \citep{Iodice_et_al_19}.
However, FCC148 hosts a Type~II break unassociated with a bar, which would imply a Type~II S0 fraction of $1/7 \approx 14^{+18}_{-9}$ per cent (excluding the three edge-on galaxies).
Clearly futher work comparing intermediate environments between field galaxies and galaxy clusters is warranted, which would enable the predictions from the simulations to be tested in more detail.

The EAGLE simulations qualitatively reproduce the observed trends found between disc profile types and galaxy mass, morphology and environment, meaning we can now use them investigate the origin of the lower Type~II frequencies in galaxy clusters.

\subsection{The lower incidence of Type~II profiles in galaxy clusters} \label{sec:TypeII_in_clusters}

\begin{figure}
  \includegraphics[width=84mm]{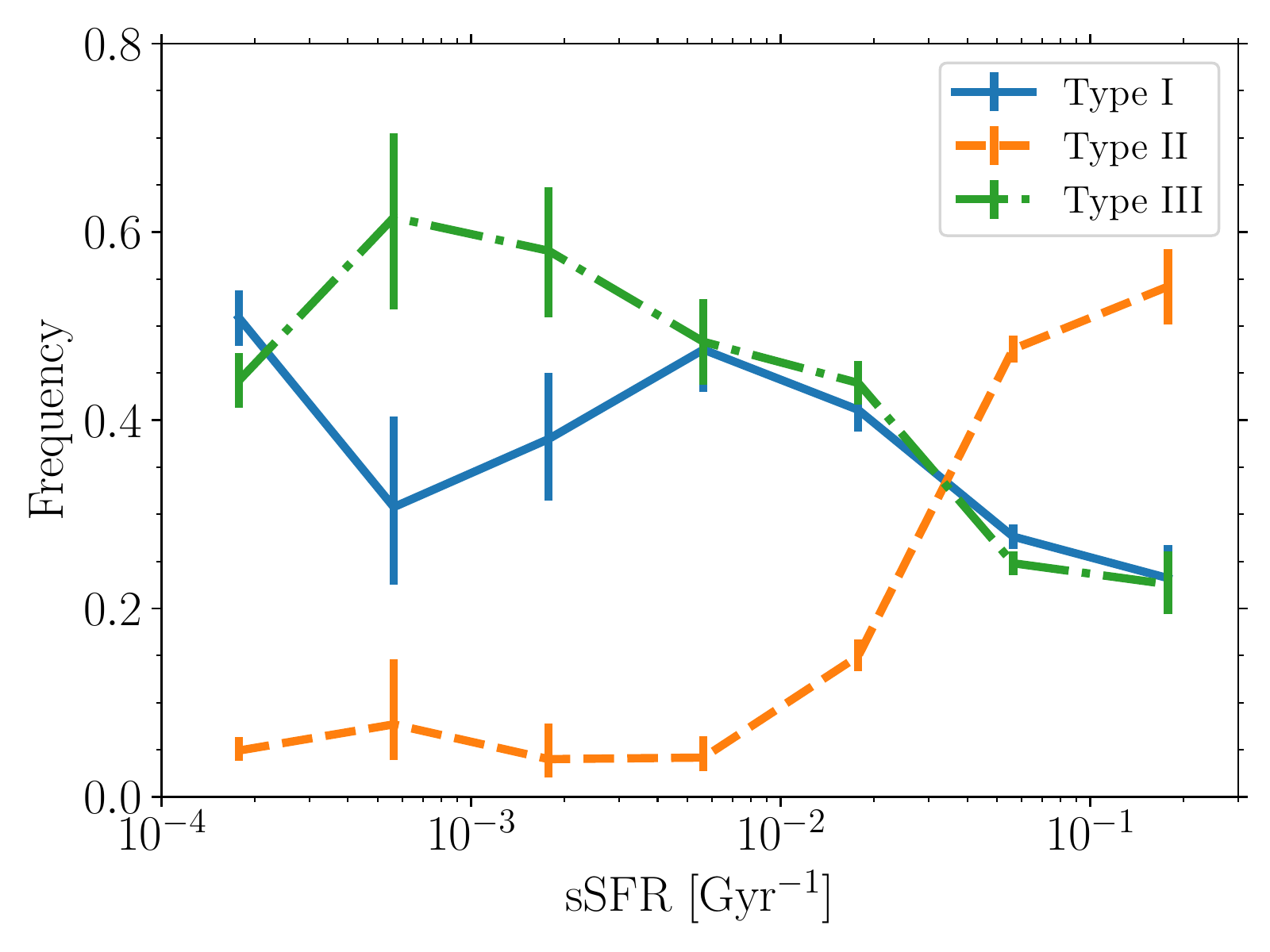}
  \caption{Frequency of disc profiles types relative to the specific star formation rates ($\mathrm{SFR}/M_\ast$) of galaxies in the EAGLE Ref-L100N1504 simulation. Galaxies with $\mathrm{sSFR} < 10^{-4} \Gyr$ are included in the lowest sSFR bin. Almost no Type~II profiles are found for galaxies with low sSFRs ($<10^{-2} \Gyr^{-1}$). Errorbars show the $1\sigma$ uncertainties from binomial statistics.}
  \label{fig:types_ssfr}
\end{figure}

\begin{figure}
  \includegraphics[width=84mm]{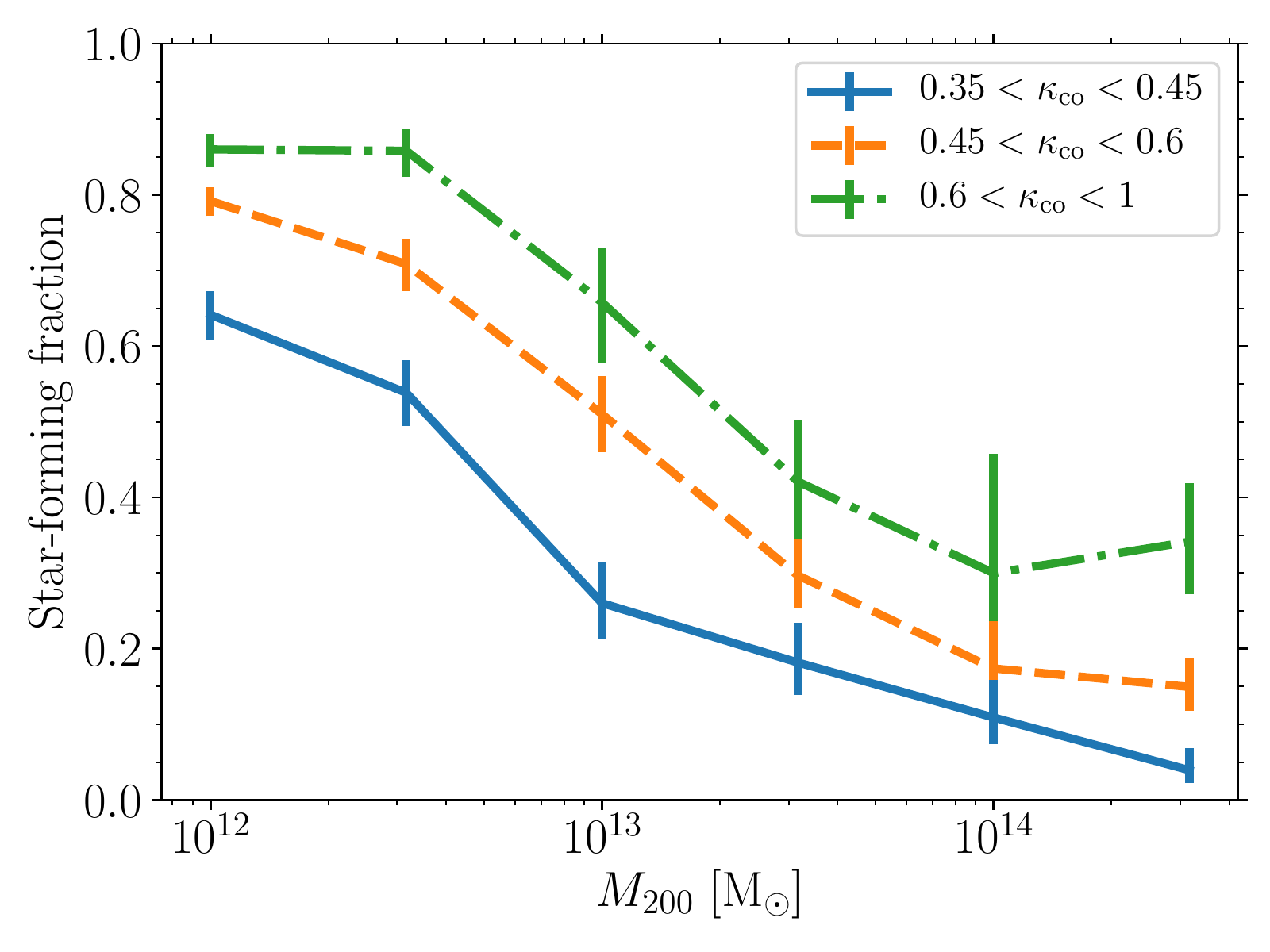}
  \caption{Fraction of EAGLE galaxies with $\mathrm{sSFR} > 0.03 \Gyr^{-1}$ (i.e. star-forming galaxies) compared with the galactic environment ($M_{200}$). The galaxy sample is divided based on the kinematic morphology $\kappaco$. The fraction of star-forming galaxies decreases consistently with increasing $M_{200}$ across all galaxy morphologies. Errorbars show the $1\sigma$ uncertainties from binomial statistics.}
  \label{fig:SFing_frac}
\end{figure}

The results of Section~\ref{sec:TypeII} show that Type~II profile depend strongly on active/recent star formation in the galaxy.
Therefore in this section we investigate the dependence of disc profiles on the SFRs of the galaxies.
When determining the SFRs of the simulated galaxies, we use the instantaneous SFR derived from the gas particle (i.e. the sum of all instantaneous SFRs for each star formation-eligible gas particle in the galaxy).

In Fig.~\ref{fig:types_ssfr} we show the frequency of disc profile types as a function of the specific star formation rate ($\mathrm{sSFR} = \mathrm{SFR}/M_\ast$) of the galaxies. 
There is a clear relationship with sSFR for Type~II profile galaxies: $48 \pm 1$ per cent of galaxies on the main star-forming sequence ($\mathrm{sSFR} > 10^{-1.5} \Gyr^{-1}$) have Type~II profiles, while only $5 \pm 1$ per cent of quiescent galaxies ($\mathrm{sSFR} \lesssim 10^{-2.5} \Gyr^{-1}$) have Type~II profiles. Instead, non-star forming galaxies are divided into Type~I and III profiles in roughly equal manner ($47 \pm 3$ per cent).

The relationship between Type~II profiles and sSFR is easily understood from the findings in Section~\ref{sec:TypeII}.
The majority of Type~II disc profiles in the EAGLE simulations are a result of truncated star-forming discs, which drive radial changes in the stellar mass-to-light ratios that cause breaks in the surface brightness profiles.
Thus, galaxies without on-going or recent star formation generally cannot have Type~II profiles.
The relationship between Type~II profiles and sSFR also explains the increasing fraction of Type~II galaxies with Hubble type \citep[Fig.~\ref{fig:types_Kappa};][]{Pohlen_and_Trujillo_06, Gutierrez_et_al_11, Laine_et_al_16, Tang_et_al_20}, given the relationship between Hubble type and SFR \citep[e.g.][]{Kennicutt_98}.

This finding thus provides a natural explanation for the lack of Type~II S0 galaxies \citep{Erwin_et_al_12} and the suppressed fraction of Type~II disc galaxies \citep[which decrease from a fraction of 50-60 percent to 34 percent,][]{Roediger_et_al_12} in the Virgo cluster, given that galaxies in denser environments show reduced sSFRs \citep[e.g.][]{Lewis_et_al_02, Kauffmann_et_al_04, Darvish_et_al_16}.
In Fig.~\ref{fig:SFing_frac} we show the fraction of star-forming galaxies (with $\mathrm{sSFR} > 10^{-1.5} \Gyr^{-1}$)\footnote{At stellar masses $M_\ast \sim 10^{10} \Msun$, the limit of $\mathrm{sSFR} > 10^{-1.5} \Gyr^{-1}$ encompasses approximately 90 per cent of `star-forming' EAGLE galaxies \citep{Furlong_et_al_15}.} as a function of the group mass ($M_{200}$) and separated by the morphology ($\kappaco$) of the galaxies.
The kinematic morphology $\kappaco$ and SFR are roughly correlated, such that high-$\kappaco$ galaxies tend to be blue and star forming, while low-$\kappaco$ galaxies tend to be red and quenched \citep{Correa_et_al_17, Correa_et_al_19}.
In `field' galaxies ($M_{200} \sim 10^{12} \Msun$) the majority of galaxies are star forming. This fraction decreases with increasing $M_{200}$, such that nearly all S0-like galaxies ($0.35 < \kappaco < 0.45$) and over half of very disc dominated galaxies ($\kappaco > 0.6$) in galaxy clusters ($M_{200} > 10^{14} \Msun$) are non-star forming. 
Combined with the very low fraction of Type~II galaxies with $\mathrm{sSFR} < 10^{-1.5} \Gyr^{-1}$ (Fig.~\ref{fig:types_ssfr}), this result explains the absence of Type~II S0 galaxies \citep{Erwin_et_al_12} and the lower fraction of Type~II disc galaxies \citep{Roediger_et_al_12} in the Virgo cluster.

Conversely, our results suggest a connection to recent star formation for the $28^{+7}_{-6}$ per cent of field S0 galaxies with Type~II profiles \citep{Erwin_et_al_12}.
This may be problematic, given S0 galaxies are typically not found on the star-forming main sequence \citep[e.g.][]{Bait_et_al_17, Mishra_et_al_19}.
Comparing the galaxy catalogues of \citet{Bait_et_al_17} and \citet{Mishra_et_al_19}, and limiting to galaxies with stellar masses $10^{10} < M_\ast / \mathrm{M}_{\sun} < 10^{10.5}$ (the most relevant comparison for the volume-limit EAGLE galaxy sample), shows that in field environments (galaxy number density $\Sigma < 10^{-0.5} \Mpc^{-2}$) only 25-35 per cent of S0 galaxies are star forming\footnote{The \citet{Bait_et_al_17} catalogue, which derives SFRs by modelling the spectral energy distribution, shows a higher fraction of star-forming galaxies than the catalogue from \citet{Mishra_et_al_19}, which use NUV$-$r colours to categorise the galaxies.}, which would imply that essentially all star-forming S0 galaxies must be Type~II.
Fig~\ref{fig:types_ssfr} suggests that, at maximum, around 50 per cent of star-forming galaxies would be expected to have Type~II profiles, indicating some tension between the simulation and observation results unless the fraction of star-forming S0 galaxies with Type~II profiles is far higher, or `green valley' galaxies also contribute to the Type~II S0 population \citep[i.e. galaxies with SFRs below the main star-forming sequence but which are not completely quenched of star formation; 20-30 per cent of the S0 population are green valley galaxies,][]{Bait_et_al_17}.

\citet{Erwin_et_al_08} found that many of the Type~II S0 galaxies have a break that coincides with an outer ring.
Such structures are common in lenticular galaxies, where around 40 per cent have outer rings \citep{Comeron_et_al_14}.
Some outer rings, and the corresponding Type~II breaks, may be related to the dynamics of bars \citep[in particular the outer Lindblad resonance,][]{Pohlen_and_Trujillo_06, Erwin_et_al_08, Munoz-Mateos_et_al_13}, though rings also occur in non-barred galaxies which might be caused by tidal interactions \citep[for a review on the formation of rings in galaxies, see][]{Buta_and_Combes_96}.
In many cases the outer rings of galaxies are UV-bright (over 50 per cent in S0 galaxies), indicating ongoing star formation \citep{Kostiuk_and_Silchenko_15, Proshina_et_al_19}.
Whether bars and/or rings in S0 galaxies result in a higher proportion of star-forming galaxies with Type~II profiles, or if some other process is at work, is unclear, and we intend to follow up this topic in future work.

\subsection{What were the disc types of S0 galaxies prior to cluster infall?} \label{sec:preinfall}

\begin{figure}
  \includegraphics[width=84mm]{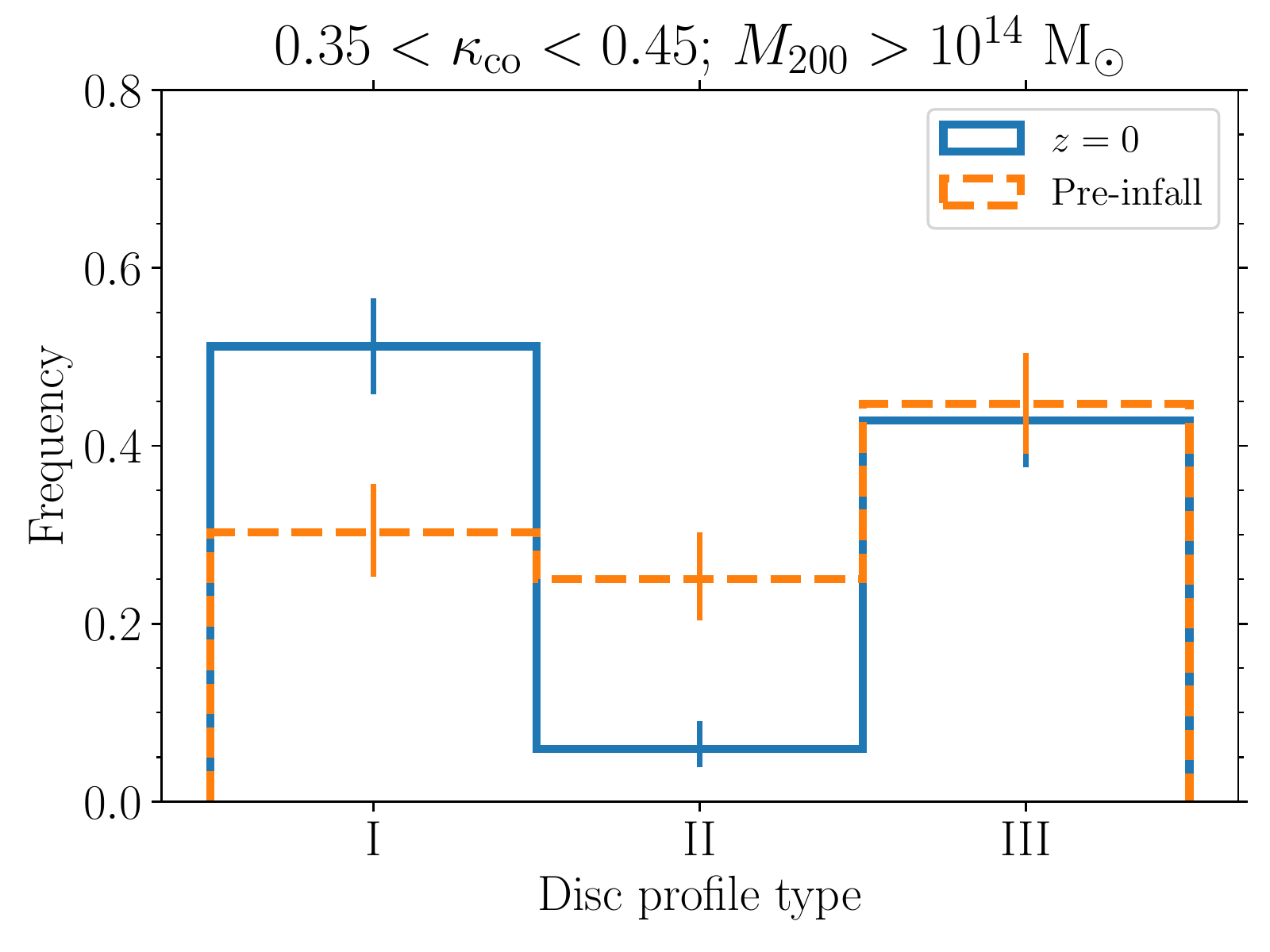}
  \caption{The frequency of disc profile types for S0-like galaxies ($0.35 < \kappaco < 0.45$) in galaxy clusters ($M_{200} > 10^{14} \Msun$) at the present day ($z=0$) and prior to becoming satellite galaxies (pre-infall). Errorbars show the $1\sigma$ uncertainties from binomial statistics.}
  \label{fig:preinfall_types}
\end{figure}

The absence of Type~II S0 galaxies in galaxy clusters leads to the question of whether the profiles were transformed within the galaxy clusters, or if Type~II profiles were already absent prior to the galaxies entering clusters.
For galaxies with $0.35 < \kappaco < 0.45$ (S0-like morphologies) and in galaxy clusters with $M_{200} > 10^{14} \Msun$ at $z=0$, we classified their $r$-band surface brightness profiles at the snapshot prior to them becoming satellite galaxies (i.e. the last snapshot at which the galaxies were considered central subhaloes in their FOF groups).

We compare the frequencies of the pre-infall disc types with the $z=0$ disc types for these galaxies in Fig.\ref{fig:preinfall_types}.
Overall, the frequency of pre-infall disc types ($30 \pm 5$, $25 \pm 5$ and $45 \pm 6$ per cent for Types~I, II and III, respectively) is relatively similar to the frequency of present day S0-like disc types for field galaxies ($37 \pm 2$, $22 \pm 2$ and $41 \pm 2$ per cent; Fig.~\ref{fig:types_Kappa}).
The fraction of Type~III galaxies is similar both before infall ($45 \pm 6$ per cent) and at $z=0$ ($43 \pm 5$ per cent), indicating anti-truncated profiles are largely unaffected by the cluster environment. 

However, the fraction of Type~II galaxies decreases from $25 \pm 5$ per cent pre-infall to $6^{+3}_{-2}$ per cent at $z=0$.
The difference is made up by an increase in the Type~I fraction at $z=0$, implying that the young stellar populations of Type~II galaxies simply fade to become Type~I galaxies upon cessation of star formation within the cluster environment (e.g. through ram pressure stripping, \citealt{Gunn_and_Gott_72}, and/or starvation, \citealt{Larson_et_al_80}). 
This finding is in good agreement with observations of Virgo cluster S0 galaxies, where the lack of Type~II profiles is similarly made up by an increased fraction of Type~I profiles relative to field galaxies \citep{Erwin_et_al_12}.

\section{Summary} \label{sec:summary}

In this work, we investigated the origin of disc profile breaks and their dependence on galaxy properties and galactic environment in the EAGLE simulations.
Following observational works, we classified $r$-band surface brightness profiles of the simulated galaxies into exponential (Type~I), truncated (Type~II) and anti-truncated (Type~III) discs.

In Section~\ref{sec:profiles} we investigated the origins of the disc profile breaks.
We found, in agreement with previous observational and theoretical work \citep{Sanchez-Blazquez_et_al_09}, that Type~II (truncated) discs predominantly result from truncated star-forming discs (Section~\ref{sec:TypeII}). 
The truncated star-forming discs cause radial gradients in the stellar populations, which in turn (through $M/L$ variations), drive breaks in the surface brightness profiles that are not necessarily found in the underlying mass profiles \citep[see also][]{Bakos_et_al_08, Sanchez-Blazquez_et_al_09, Tang_et_al_20}.
In the EAGLE model, the truncated star-forming discs result from density thresholds for star formation and occur when the total gas density falls below $\sim10^{6.5} \Msun \kpc^{-2}$ \citep[see][]{Schaye_04, Schaye_and_Dalla_Vecchia_08}.
Upon entering galaxy groups and clusters, and following a cessation in star formation, in most cases the young stellar populations of Type~II galaxies simply fade, leaving behind Type~I galaxies (Section~\ref{sec:preinfall}).
This agrees well with observational evidence suggesting S0 galaxies in large groups are predominantly faded spiral galaxies \citep{Deeley_et_al_20} and simulations showing S0-progenitor galaxies lose their gas soon after group/cluster infall (Deeley et al., MNRAS, submitted).

In contrast, we found that Type~III (anti-truncated) discs have a more varied origin (Section~\ref{sec:TypeIII}).
The profile breaks are also present in the underlying mass profiles, thus are not driven by stellar population gradients like for Type~II discs.
In the simulations we found examples of anti-truncations created through galaxy mergers (both major and minor mergers, extended star-forming discs and the late build-up of a steeper inner disc.
The increasing importance of mergers for stellar mass growth in massive galaxies \citep[e.g.][]{Rodriguez-Gomez_et_al_16, Qu_et_al_17} suggests that the increasing fraction of Type~III galaxies with galaxy mass (Fig.~\ref{fig:types_Mstar}) is driven by galaxy mergers.

In Section~\ref{sec:frequencies} we investigated the origin of the lower incidence (relative to field galaxies) of Type~II galaxies in galaxy clusters \citep{Erwin_et_al_12, Roediger_et_al_12, Raj_et_al_19}.
In agreement with observational studies, we found in the simulations that almost no S0-like galaxies in galaxy clusters have Type~II profiles, while the fraction of Type~II profiles for disc-dominated galaxies also significantly decreases (Fig.~\ref{fig:types_Kappa}).
We found that this decrease in Type~II fraction can be directly attributed to the lower SFRs of galaxies residing in clusters: the frequency of Type~II profiles is a strong function of the galaxies' specific SFR (Fig.~\ref{fig:types_ssfr}) and the fraction of star-forming galaxies is a strong function of group/cluster mass (Fig.~\ref{fig:SFing_frac}).
Therefore, low SFR galaxies in galaxy clusters generally do not host Type~II profiles, unlike their star-forming morphological analogues in the field environment.

\section*{Acknowledgements}

We thank Michael Drinkwater and Camila Correa for helpful comments and discussion, and the referee for a very constructive report which improved the paper.
This research was supported by the Australian government through the Australian Research Council’s Discovery Projects funding scheme (DP200102574).
This work used the DiRAC Data Centric system at Durham University, operated by the Institute for Computational Cosmology on behalf of the STFC DiRAC HPC Facility (\url{www.dirac.ac.uk}). This equipment was funded by BIS National E-infrastructure capital grant ST/K00042X/1, STFC capital grants ST/H008519/1 and ST/K00087X/1, STFC DiRAC Operations grant ST/K003267/1 and Durham University. DiRAC is part of the National E-Infrastructure.

\section*{Data Availability}

All data (including galaxy catalogues, merger trees and particle data) from the EAGLE simulations is publicly available \citep{McAlpine_et_al_16} at \href{http://www.eaglesim.org/database.php}{http://www.eaglesim.org/database.php}.



\bibliographystyle{mnras}
\bibliography{bibliography} 






\bsp	
\label{lastpage}
\end{document}